\documentclass[a4paper,12pt]{article}
\pdfoutput=1

\usepackage{jheppub,float}
\usepackage{amsmath,amssymb,amsfonts}
\usepackage{graphicx}
\usepackage{cleveref}
\usepackage{hyperref}
\usepackage{srcltx}
\usepackage{bm}
\usepackage{mathdots}
\usepackage{color}
\usepackage{dsfont}
\usepackage{slashed}
\usepackage{physics}
\usepackage{subfigure}
\usepackage{tikz}
\usepackage[normalem]{ulem}

\newcommand{\al}[1]{\begin{align} #1 \end{align} }
\newcommand{\prn}[1]{ \left(  #1 \right) }
\newcommand{\avg}[1]{\langle #1 \rangle}
\newcommand\beq{\begin{eqnarray}}
\newcommand\eeq{\end{eqnarray}}
\newcommand{\TR}{T_\text{R}}
\newcommand{\Tmax}{T_\text{max}}
\newcommand{\Mpl}{M_\text{Pl}}

\definecolor{sky}{rgb}{0., 0.5883488108126352, 0.9445016153846155}

\preprint{LCTP-22-13 \\ \vspace{-8mm}\hfill MITP-22-085}
\title{Freezing-in hadrophilic dark matter \\ at low reheating temperatures}
\author[a]{Prudhvi N.~Bhattiprolu,} \emailAdd{prudhvib@umich.edu}
\author[b]{Gilly Elor,}\emailAdd{gelor@uni-mainz.de}
\author[a]{Robert McGehee,}\emailAdd{rmcgehee@umich.edu}
\author[a]{and Aaron Pierce}\emailAdd{atpierce@umich.edu}
\affiliation[a]{Leinweber Center for Theoretical Physics, Department of Physics,\\ University of Michigan, Ann Arbor, MI 48109, USA}
\affiliation[b]{\sl PRISMA$^+$ Cluster of Excellence \& Mainz Institute for Theoretical Physics\\
Johannes Gutenberg University, 55099 Mainz, Germany\\[2mm]}

\abstract{If the reheating temperature at the end of inflation is low, of order 10 MeV, then dark matter produced through ultraviolet freeze-in has a large direct detection cross section. We study such a scenario in which dark matter is hadrophilic. This leads to dark matter-nucleon scattering cross sections of interest for near-future experiments for dark matter masses in the range of 100 keV-100 MeV. We explore how these predictions vary if reheating is non-instantaneous.} 

\begin{document}
\maketitle
\flushbottom
\setcounter{page}{2}
\newpage

\section{Introduction}
The search for weakly interacting massive particles is mature, and direct detection experiments have placed strong bounds on this dark matter paradigm. However, flagship experiments such as Xenon-nT~\cite{XENON:2020kmp}, LZ~\cite{Aalbers:2022fxq}, and PandaX~\cite{PandaX-4T:2021bab} have weakened sensitivity to dark matter with masses less than $\mathcal{O}(10 \text{ GeV})$ due to their energy thresholds. But dark matter with masses as low as $\mathcal{O}(10 \text{ keV})$ produced from the thermal bath of Standard Model (SM) particles can be consistent with bounds from large scale structure and Lyman-$\alpha$ measurements~\cite{Ballesteros:2020adh}. This has motivated the development of novel detector materials and technologies that can probe this interesting keV to GeV window (see \emph{e.g.}~\cite{Knapen:2016cue,Budnik:2017sbu,Knapen:2017xzo,Knapen:2017ekk,Griffin:2018bjn,Kurinsky:2019pgb,Essig:2019kfe,Trickle:2019nya,Griffin:2019mvc,Campbell-Deem:2019hdx,Griffin:2020lgd,Coskuner:2021qxo,Campbell-Deem:2022fqm}). Some proposals hope to probe dark matter-nucleon cross sections as small as $10^{-45} - 10^{-43} \text{ cm}^2$ after only $\sim 1 \text{ kg}\cdot\text{yr}$ exposure.

However, the number of future detection technologies may exceed the number of detectable sub-GeV dark matter candidates with concrete cosmological histories, especially those which preferentially couple to hadrons. The few sub-GeV hadrophilic dark matter candidates include asymmetric dark matter~\cite{Knapen:2017xzo,Coskuner:2018are}, Co-SIMPs~\cite{Smirnov:2020zwf}, and HYPERs~\cite{Elor:2021swj}. One challenge in constructing such models is a robust lower bound of $\mathcal{O}(10 \text{ MeV})$ on the mass of dark matter particles which were in chemical equilibrium with the SM bath~\cite{An:2022sva}. These light particles would unavoidably contribute to the relativistic degrees of freedom at the time of big bang nucleosynthesis (BBN), altering the primordial abundances of light nuclei. Additionally, cross sections that are directly detectable for light dark matter often require light mediators. These light mediators can also contribute to dark radiation at the time of BBN.

One way around these difficulties is to consider other production mechanisms, such as freeze-in \cite{Hall:2009bx} wherein feeble interactions between the SM and the dark matter never result in a chemical equilibrium between the two; nevertheless, a significant dark matter density may be built up. Models utilizing this mechanism give a direct-detection benchmark for sub-GeV dark matter scattering off {\it electrons}~\cite{Essig:2011nj}, and is a target for the SENSEI experiment~\cite{SENSEI:2020dpa}. 

With these future low-threshold experiments and model-building challenges in mind, we explore a model of sub-GeV hadrophilic dark matter coupled to the SM via a scalar mediator. We find that UV freeze-in~\cite{Elahi:2014fsa} scenarios in which the dark matter is frozen-in instantaneously at a low reheating temperature $\TR$ through a higher dimension operator can provide interesting direct-detection benchmarks.\footnote{UV freeze-in dark matter in low reheating scenarios has also been considered in the context of future high-luminosity experiments such as LDMX~\cite{Berlin:2018bsc}.} Large couplings to the Standard Model are required to reproduce the relic abundance if $\TR$ is as low as possible without disturbing the successful predictions of BBN, $\TR \sim 10 \text{ MeV}$.\footnote{Generating the baryon asymmetry at such low reheating temperatures is possible~\cite{Dimopoulos:1987rk,Cline:1990bw,Aitken:2017wie,Elor:2018twp,Elor:2020tkc,Elahi:2021jia,Jaeckel:2022osh}. Alternatively, a large pre-existing baryon asymmetry may be diluted, and we do not address this further.} The predicted direct detection cross section depends on whether the maximum temperature of the SM bath during pre-heating, $\Tmax$, is close to or much greater than $\TR$. We consider in detail both instantaneous reheating and finite reheating scenarios. 

The outline of this paper is as follows. In Sec.~\ref{sec:Model}, we present a model of hadrophilic dark matter with a scalar mediator. In Sec.~\ref{sec:InstanRH}, we analyze the UV freeze-in of dark matter assuming instantaneous reheating and predict direct detection cross sections for future dark matter experiments for various low $\TR$. In Sec.~\ref{sec:FiniteRH}, we relax the assumption of instantaneous reheating, allowing $\Tmax > \TR$. Freeze-in processes with initial-state pions are especially sensitive to $\Tmax$ and can significantly impact the predicted cross section. We conclude in Sec.~\ref{sec:disc}. In App.~\ref{sec:altUV}, we consider variations on the model presented in the text and discuss how the choice of model impacts the predicted direct detection cross section. In App.~\ref{sec:FiniteReheatingYield}, we derive an expression for the contribution to the dark matter yield from $\Tmax$ to $\TR$.

\section{A model of hadrophilic dark matter}
\label{sec:Model}

Our UV-complete model of hadrophilic dark matter contains a real scalar SM singlet $\phi$, TeV-scale colored vectorlike fermions $\psi$, and a fermionic dark matter SM singlet $\chi$.\footnote{Similar hadrophilic UV completions have been explored in different contexts (\emph{e.g.}~\cite{Knapen:2017xzo,Elor:2021swj}).} The Lagrangian contains the following terms
\beq
\label{eq:Lagr}
\mathcal{L}
&\supset&
 - m_\psi \overline{\psi} \psi - m_\chi \overline{\chi} \chi - \frac{1}{2} m_\phi^2 \phi^2 - y_\psi \phi \overline{\psi} \psi  - y_\chi \phi \overline{\chi} \chi,
\eeq
which respect a $\mathbf{Z}_2$ symmetry under which $\psi$ is odd.

Because $\psi$ is colored, it induces a coupling of $\phi$ to hadrons when integrated out, so  $\phi$ is a hadrophilic mediator. The rest of $\psi$'s SM charge assignment will impact the details of the phenomenology. For concreteness, we take $\psi$ to transform as $\prn{\textbf{3},\textbf{2},1/6}$, under the SM gauge group $\left (\text{SU}(3)_c, \text{SU}(2)_L, \text{U}(1)_Y \right)$, which would allow for a potential eventual embedding in a Grand Unified Theory (GUT) framework.  However, we do not assume supersymmetry nor complete GUT multiplets for $\psi$, so substantial threshold corrections would be needed to obtain unification.  For  discussion how on the choice of charges for the fermions (including the possibility of adding a full GUT multiplet) impacts the results presented here, see App.~\ref{sec:altUV}. 

As long as the $\mathbf{Z}_2$ breaking is small (or zero), $\psi$ are stable over detector lengths and we may infer
\al{
m_\psi \gtrsim 1.5 \text{ TeV} \quad \text{(95\% CL)}
\label{eq:mpsibnd}
}
from the LHC searches for long-lived bottom squarks based on ionization energy loss and time of flight by the ATLAS collaboration~\cite{ATLAS:2019gqq}. 

After $\psi$ is integrated out, $\phi$ has an effective coupling to gluons  
\al{
\mathcal{L} \supset  \frac{y_\psi \alpha_s}{6 \pi m_\psi} \phi G^a_{\mu \nu} G^{\mu \nu,a}.
}
In the low-energy theory, this translates into a $\phi$-nucleon coupling given by \cite{Gunion:1989we}
\al{
\mathcal{L} &\supset -y_n \phi \overline{n} n,
&y_n = y_\psi \frac{4 m_n}{27 m_\psi},
\label{eq:yn}
}
where $m_n$ is the mass of the neutron. Here and below, we use $m_p \simeq m_n$ where $m_p$ is the proton mass. We only consider $m_{\phi}$ much greater than the momentum transfer at direct detection experiments. Thus, the dark matter-nucleon scattering cross section relevant for direct detection is
\al{
\sigma_{\chi n} = \frac{\left ( y_n y_\chi \right )^2}{\pi} \frac{\mu_{\chi n}^2}{m_\phi^4}\,,
\label{eq:scattering_sigma}
}
where $\mu_{\chi n}$ is the dark matter-nucleon reduced mass. 

Since we will be focused on low $\TR \ll \text{GeV}$, neither the gluon coupling nor the nucleon coupling is the most relevant for freeze-in. Instead, induced couplings to pions and photons matter most. The $\phi$ coupling to pions is \cite{Gunion:1989we}
\al{
\mathcal{L} \supset  \frac{3 y_n }{m_n} \phi \prn{\frac{2}{3} \left|D^\mu \pi^+\right|^2 - m_\pi^2 \pi^+ \pi^-}.
\label{eq:phipipi}
}
$\phi$ also couples to other hadrons. These hadronic couplings, together with the coupling of $\phi$ to $\psi$ (which is charged), generate a coupling of $\phi$ to photons:
\al{
\mathcal{L}_{\phi FF} \sim \frac{17 y_n \alpha}{8 \pi m_n} \phi F_{\mu \nu} F^{\mu \nu}.
\label{eq:phiFF}
}
This coupling gets a contribution from both IR (the charged hadrons) and UV ($\psi$) physics. While the contribution to the coefficient of this operator from the UV is calculable, the IR contribution is subject to non-perturbative physics. For this IR contribution, we use the Naive Dimensional Analysis (NDA) \cite{Georgi:1986kr,Manohar:1983md} estimate $\frac{y_n \alpha}{4 \pi m_n}$. The UV contribution to this coefficient, which depends on the electroweak charge of $\psi$, is $\frac{15 y_n \alpha}{8 \pi m_n}$ for the representation chosen above. As we will see, the induced couplings in Eqs.~\eqref{eq:phipipi} and \eqref{eq:phiFF} permit dark matter to freeze-in through pion and photon annihilations. 

Last, we summarize phenomenological constraints on the Yukawa couplings of Eq.~(\ref{eq:Lagr}) as well as $m_\phi$.
For
$m_{\phi}$ lighter than the range considered below,
$y_{\chi}$ would be constrained by limits on self-interacting dark matter \cite{Randall:2008ppe,Kaplinghat:2015aga}.
Here, however, both $y_\psi$ and $y_\chi$ are only constrained by perturbativity. In our results, we set both to a maximum value of 1. The lower bound on $m_\psi$ in Eq.~\eqref{eq:mpsibnd} implies an upper-bound on $y_n$ through Eq.~\eqref{eq:yn},
\begin{equation}
y_{n} = 9.3 \times 10^{-5} \left(\frac{y_{\psi}}{1}\right)\left(\frac{1.5 \, \rm{TeV}}{m_{\psi}}\right).
\label{eq:ynbnd}
\end{equation}
There are also constraints in the $y_n$--$m_\phi$ plane arising from rare meson decays. The $\psi$'s give rise to a $\phi \overline{t} t$ coupling at two-loop order \cite{Knapen:2017xzo}:
\al{
\mathcal{L} &\supset - y_t \phi \overline{t} t,
&y_t
= y_n \frac{9 \alpha_s^2}{2 \pi^2} \frac{m_t}{m_n} \log\left( \frac{m_\psi^2}{m_t^2} \right).
}
This coupling in turn contributes to $B^+ \to K^+ \phi$ decays, $K^+ \to \pi^+ \phi$ decays, and $B-\bar{B}$ mixing. Constraints from these processes are weaker than Eq.~\eqref{eq:ynbnd}. The bounds from rare $B^+$ \cite{BaBar:2013npw} and $K^+$ \cite{NA62:2021zjw} decays have been discussed previously \cite{Knapen:2017xzo}. We have checked $B-\bar{B}$ mixing constraints \cite{Charles:2020dfl} do not limit the allowed parameter space, and we do not discuss them further.

\section{UV freeze-in with instantaneous reheating}
\label{sec:InstanRH}

Following the era of inflation, reheating could occur rapidly. If it does, the maximum temperature of the SM bath after inflation $\Tmax$, could coincide with the temperature that marks the beginning of the radiation domination era $\TR$. One cosmological history which could achieve $\Tmax \simeq \TR$ is a multi-field inflation scenario, perhaps accomplishing reheating via a parametric resonance, see e.g. \cite{Kofman:1997pt}. In this section, we assume that $\TR \simeq \Tmax$. In the next section, we discuss the implications of relaxing this assumption, allowing $\Tmax > \TR$.

\subsection{Freeze-in calculations}
\label{sec:FI1}

In this subsection, we calculate the UV freeze-in of dark matter from SM particle annihilations for the case of instantaneous reheating. Since $\phi$ is hadrophilic, and we focus on low reheating temperatures, $5 \text{ MeV} \lesssim \TR \lesssim 20 \text{ MeV}$, freeze-in proceeds through photon and---perhaps less obviously---pion annihilations. For even though the abundance of pions are Boltzmann suppressed at these temperatures, they can have a significant impact. The lower bound on the reheating temperature is set by the requirement that predictions of BBN remain undisturbed~\cite{Hannestad:2004px,deSalas:2015glj}. The upper bound, as we will show below, corresponds roughly to the $\TR$ at which predicted direct detection cross sections are too small to be observable at future experiments. Note the $\mathcal{O}(\text{TeV})\; \psi$ are not produced in the early universe due to the low $\TR$. 

We first calculate the dark matter produced from photon annihilations, $\gamma \gamma \to \chi \overline{\chi}$. The UV freeze-in proceeds through the dimension-7 operator (see discussion surrounding Eq.~\eqref{eq:phiFF}),
\beq
\label{eq:dim7}
    \mathcal{L} &\supset& 
    \frac{y_\chi}{m_\phi^2} \frac{17 y_n \alpha}{8 \pi m_n} \overline{\chi} \chi F_{\mu \nu} F^{\mu \nu}.
\eeq
Ignoring the possible contribution from initial-state pions for the moment, the number-density Boltzmann equation governing dark matter freeze-in is \cite{Kolb:1990vq}
\al{
\dot{n}_\chi + 3H n_\chi \approx  \avg{\sigma_{\chi \overline{\chi} \to \gamma \gamma} v} \prn{n_\chi^\text{eq}}^2, \nonumber \\
\avg{\sigma_{\chi \overline{\chi} \to \gamma \gamma} v} \prn{n_\chi^\text{eq}}^2 = \frac{1}{2!} \frac{g_\chi^2 g_\gamma^2 T}{2^9 \pi^5} \int_{4 m_\chi^2}^\infty ds \sqrt{s-4m_\chi^2} K_1 \prn{\sqrt{s}/T} \overline{\left| \mathcal{M} \right|^2},
\label{eq:boltz}
}
where $\overline{\left| \mathcal{M} \right|^2}$ is the fully averaged matrix element squared for the process. The $1/2!$ appears due to the 2 identical photons. In the second line, $g_\chi$ and $g_\gamma$ are the degrees of freedom of dark matter and photons. The phase space has been simplified to a single integral since $\overline{\left| \mathcal{M} \right|^2}$ is only a function of $s$ (see \emph{e.g.}~\cite{Elahi:2014fsa,Hochberg:2018rjs}). This assumes Maxwell-Boltzmann distributions for thermal $\gamma$'s and $\chi$'s and that $\chi$'s are always out of equilibrium.

Proceeding in a similar vein, we can  calculate the UV freeze-in contribution coming from pion annihilations $\pi^+ \pi^- \to \bar{\chi} \chi$ due to the couplings in Eq.~\eqref{eq:phipipi}. Adding both contributions and integrating over temperatures, we obtain the yield of dark matter, $Y_\text{DM}=n_\text{DM}/s$, the ratio of the dark matter number density to the SM bath entropy. The total yield obtained from pions and photons is
\al{
Y_\text{DM}= \frac{y_n^2 y_\chi^2}{m_\phi^4 m_n^2} \frac{135 \sqrt{10} \Mpl}{(2\pi)^8} \sum_{j=\gamma,\pi^\pm} \kappa_j^2 \int_{0}^{\TR} \frac{dT}{T^5} \frac{I(m_j)}{g_{s,\ast}(T) \sqrt{g_\ast(T)}},
\label{eq:Yinstan}
}
where
\beq
\kappa_j &=& 
\begin{cases}
1 \text{ for } j = \pi^\pm,\\
\frac{17\alpha}{4 \pi}
\text{ for } j = \gamma,
\end{cases}
\label{eq:kappa_yield_repQ}
\eeq
and
\begin{alignat}{1}
\label{eq:Idef}
I(m)
&=
\int_{{\rm max}(4 m^2, 4 m_\chi^2)}^{\infty} ds \frac{\left( s +m^2\right)^{2}}{\left( \frac{s}{m_\phi^2} - 1\right)^2 + \frac{\Gamma_\phi^2}{m_\phi^2}}
\left( s - 4 m_\chi^2\right)^{3/2} \sqrt{1 - \frac{4 m^2}{s}} 
K_1 \left( \frac{\sqrt{s}}{T} \right).
\end{alignat}
Above, $\Mpl=2.4\times 10^{18}$ GeV is the reduced Planck mass, $g_{(s,)\ast}$ are the (entropic) relativistic degrees of freedom, and $\Gamma_\phi$ is the total width of $\phi$ which in our case almost always decays to $\chi \overline{\chi}$. The presence of this width as well as $s/m_\phi^2$ in the denominator of Eq.~\eqref{eq:Yinstan} is due to our use of the more general Lagrangian in Eq.~\eqref{eq:phiFF} instead of the approximation in Eq.~\eqref{eq:dim7}. For instantaneous reheating, $g_{(s,)\ast}$ are not changing abruptly during freeze-in and the relativistic degrees of freedom are just evaluated at $\TR$.\footnote{We have also verified that $dg_{s,\ast}/dT$ is sufficiently negligible in both the instantaneous and non-instantaneous reheating scenarios.} To explain the observed dark matter, we require 
\al{
m_\chi Y_\text{DM} = \frac{\Omega_\text{DM} \rho_{crit}}{s_0} = 4.37 \times 10^{-10} \text{ GeV},
\label{eq:DMabund}
}
where $\Omega_\text{DM}$ is the energy fraction of dark matter, $\rho_{crit}$ is the critical energy density, and $s_0$ is the entropy of the photon bath today \cite{Workman:2022ynf}. 

\begin{figure}[t!]
    \begin{center}
        \includegraphics[width=0.95\textwidth]{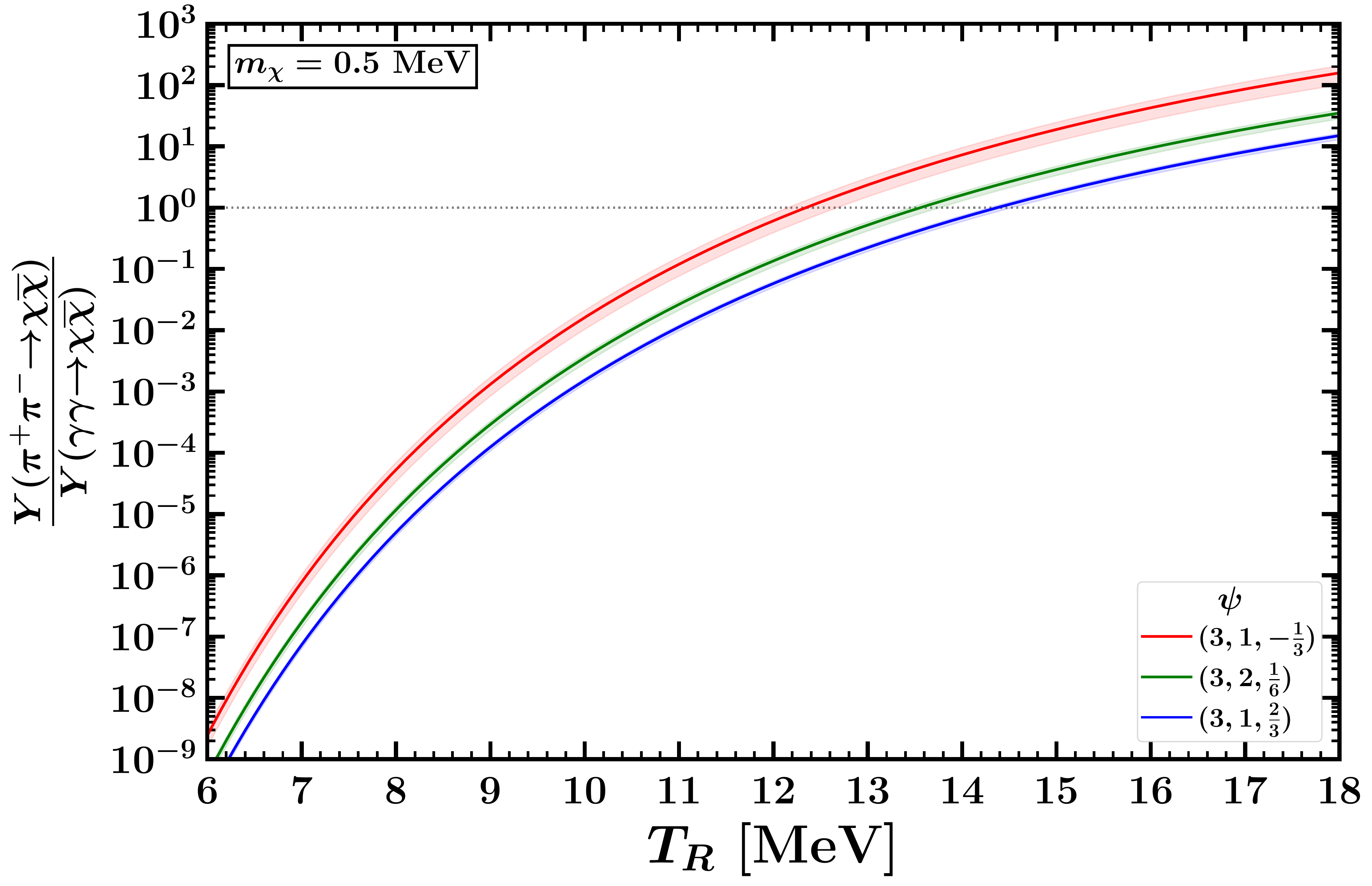}
    \end{center}
    \caption{The ratio of dark matter produced from pions to dark matter from photons as a function of the reheating temperature for $m_\chi = 0.5$ MeV. Different curves correspond to different representations of $\psi$ under the SM gauge group $\left(\text{SU}(3)_c,\text{SU}(2)_L,\text{U}(1)_Y \right)$. The quark doublet-like representation on which we focus in the main text is shown in green. 
    \label{fig:InPi2PhYvsTR}}
\end{figure}
For the low reheating temperatures we consider, one might suspect that photon annihilations are the dominant process for dark matter freeze-in. For $\TR \lesssim 13 \text{ MeV}$, this is the case, as shown in Fig.~\ref{fig:InPi2PhYvsTR}. This figure displays the yield of dark matter frozen-in via pions divided by the corresponding yield frozen-in via photons as a function of the reheating temperature, i.e., the ratio of the two terms in the sum of Eq.~\eqref{eq:Yinstan}. The different colored curves correspond to different possible SM charge assignments for $\psi$. The quark doublet-like representation we have emphasized is displayed in green. The band around the curves corresponds to varying the IR contribution to the photon coupling around the NDA estimate of Eq.~\eqref{eq:phiFF} by a factor of 2. As expected, the width of the band is  smaller for those representations that receive larger contributions to $\phi FF$ from integrating out $\psi$ in the UV. For now we continue to concentrate on $\psi \sim \prn{\textbf{3},\textbf{2},1/6}$, and we postpone discussion of other representations to App.~\ref{sec:altUV}. 

Returning to Fig.~\ref{fig:InPi2PhYvsTR}, we note that for temperatures $\TR \gtrsim 13 \text{ MeV}$, pion annihilations are the dominant freeze-in channel.\footnote{We have checked that $\pi^\pm \gamma \to \pi^\pm \bar{\chi} \chi$ never matters: it only produces more dark matter than $\pi^+ \pi^- \to \bar{\chi} \chi$ for $\TR$ low enough that photon annihilations dominate.} This is despite the exponential penalty required to find two pions in the relatively cold thermal bath. The presence of this additional channel   reduces the predicted direct-detection cross sections since it requires smaller couplings than would naively be predicted if only photon annihilations were taken into account. It also foreshadows the importance of the assumption $\Tmax = \TR$ that we have made thus far. Allowing for $\Tmax > \TR$ can substantially increase the yield from pions, as we will see in the next section.

One final comment: electrons will also be in the SM bath at low $\TR$ and in principle could contribute to the freeze-in of dark matter. However, relative to the yield from photon annihilations in Eq.~\eqref{eq:Yinstan}, we expect this yield to be suppressed by at least $\alpha^2 m_{e}^2/\TR^2$ and is therefore negligible.

\subsection{Results}
\label{sec:results1}

\begin{figure}[t!]
    \begin{center}
        \includegraphics[width=\textwidth]{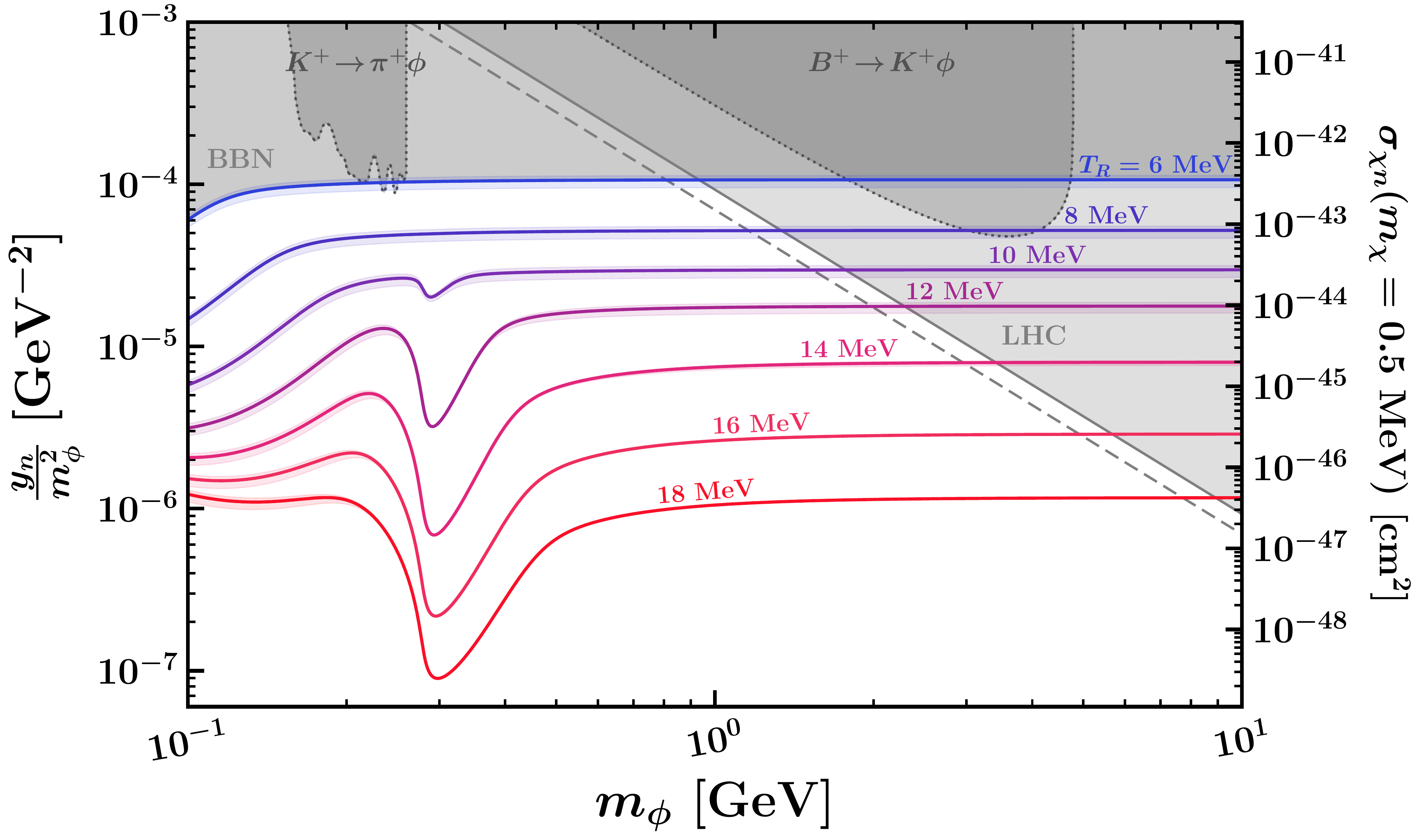}
    \end{center}
    \caption{Predictions for $y_n/m_{\phi}^2$ as a function of $m_\phi$ along with constraints in gray for $y_\psi = y_\chi = 1$. The LHC constraint is obtained from Eq.~(\ref{eq:ynbnd}) with $m_\psi \gtrsim 1.5$ ($2$) TeV for shaded region (long dashed line) for current (possible future) constraint. The colored contours yield the correct relic abundance for $m_\chi = 0.5 \text{ MeV}$ for various reheating temperatures. The right vertical axis is the scattering cross section for $m_\chi = 0.5 \text{ MeV}$. 
    \label{fig:ynovrmphi2}}
\end{figure}

For fixed $m_\chi$ and $\TR$, explaining the dark matter abundance in Eq.~\eqref{eq:DMabund} predicts a value of $y_n/m_\phi^2$. We plot this requisite value as a function of $m_\phi$ in Fig.~\ref{fig:ynovrmphi2} for a benchmark case of $\left(m_\chi = 0.5 \text{ MeV}, y_\psi = 1, y_\chi = 1\right)$ and various reheating temperatures. The shaded bands surrounding these curves (which are noticeable only for lower reheating temperatures) correspond to varying the NDA estimate for the photon coupling in Eq.~\eqref{eq:phiFF} by a factor of 2. Uncertainty in the photon coupling due to NDA estimates changes the predicted $\sigma_{\chi n}$ by roughly 15\% for lower reheating temperatures, and this effect gets smaller as the contribution from the pions outstrips that of the photons at higher temperatures. We will not show this uncertainty in the rest of our results. 

The constraints on $y_n$ and $m_\phi$ discussed in Sec.~\ref{sec:Model} are shown in shades of gray. The light gray shaded region is ruled out by current direct searches for $\psi$ from the LHC; see Eq.~\eqref{eq:ynbnd}. The dashed gray line assumes a possible future bound from the LHC of $m_\psi \gtrsim 2 \text{ TeV}$. We also show the approximate lower bound on $\TR$ coming from BBN in light gray. The darker gray regions are constrained by searches for rare $B$ and $K$ decays.  These constraints are subdominant except for a narrow range of $\phi$ masses near $m_\phi \sim 200 \text{ MeV}$. 

A noteworthy feature of this plot is the resonance around $m_\phi \sim 2 m_{\pi^+}$. This resonance causes pion annihilations to become more productive and diminishes the requisite nucleon coupling and cross section. Because one point of interest is to find benchmarks for future direct detection efforts, to achieve larger cross sections and simplify our analysis, we require $m_{\phi} \geq 1 \text{ GeV}$ in the rest of this section. For these heavier $\phi$'s, the requisite value of $y_n/m_\phi^2$ is roughly independent of $m_\phi$ (as evident in Fig.~\ref{fig:ynovrmphi2}). In our numerical computations, we do not assume UV freeze-in and calculate the freeze-in with the full $\phi$ propagator, allowing for its on-shell production and subsequent decay to pairs of dark matter. However, once we impose this lower bound on $m_\phi$, the effective operator picture of Eq.~\eqref{eq:dim7} is valid~\cite{Frangipane:2021rtf} and allows our results to conform to some of the usual intuition from simpler UV freeze-in scenarios.

\begin{figure}[t!]
    \begin{center}
        \includegraphics[width=0.95\textwidth]{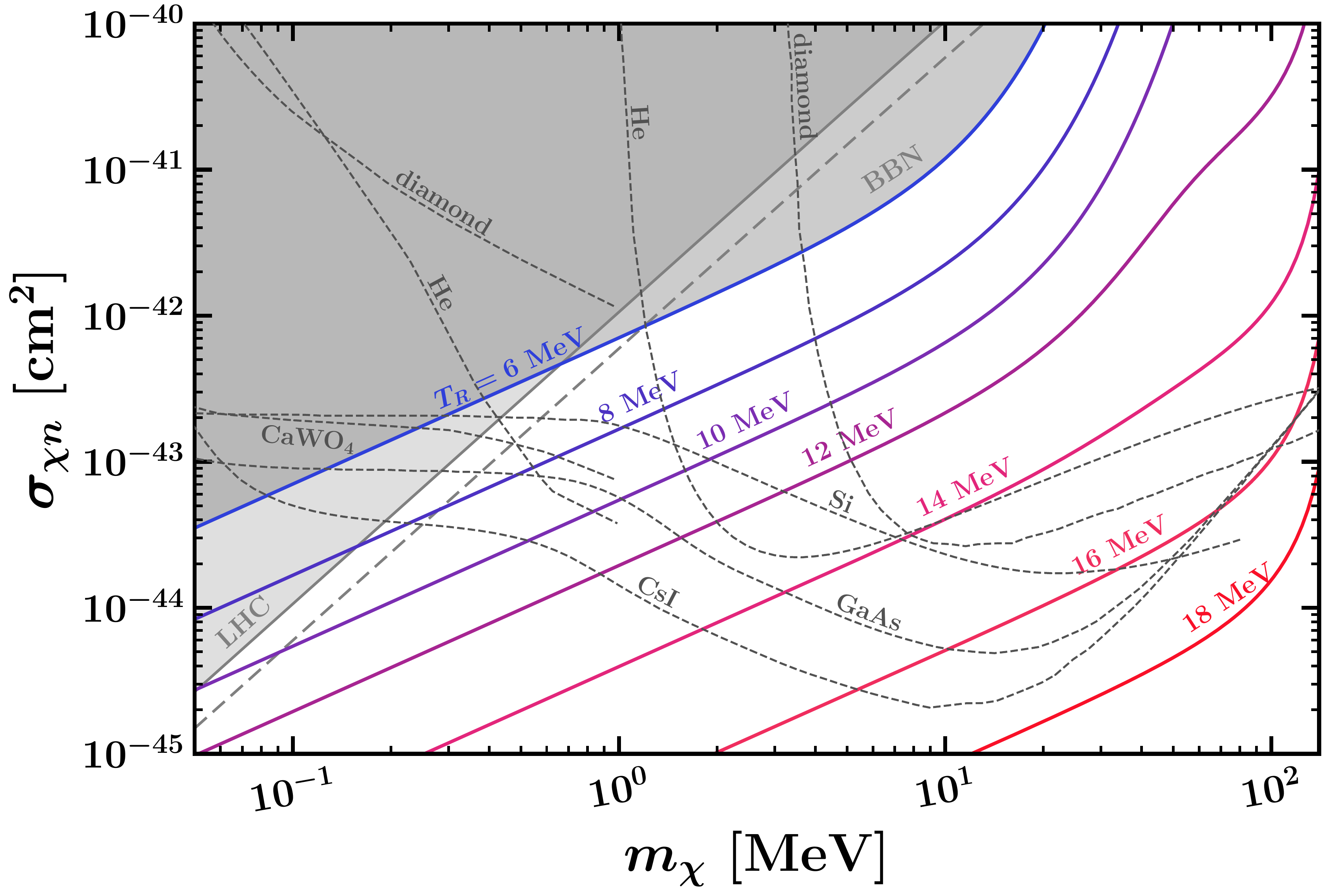}
    \end{center}
    \caption{The dark matter-nucleon scattering cross section as a function of $m_\chi$ which yields the correct relic abundance for $y_\psi = y_\chi = 1$. The colored contours correspond to different reheating temperatures. The gray shaded regions are excluded by the LHC and BBN, while the dashed gray lines show the projected sensitivities of future experiments \cite{Knapen:2016cue,Kurinsky:2019pgb,Trickle:2019nya,Griffin:2019mvc,Campbell-Deem:2019hdx,Coskuner:2021qxo,Campbell-Deem:2022fqm}.}
    \label{fig:sigmachin}
\end{figure}

The right vertical axis of Fig.~\ref{fig:ynovrmphi2}  shows that low reheating temperatures may predict detectable cross sections. Going to large $\TR$ rapidly decreases the cross section, even more quickly than if only photon annihilations were considered. Despite their exponentially small abundance in the thermal bath, pions quickly overtake photons to become the dominant driver of freeze-in for $\TR \gtrsim 14 \text{ MeV}$ (see Fig.~\ref{fig:InPi2PhYvsTR}). Nevertheless, it is informative to first consider the limiting case in which $\TR \lesssim 12 \text{ MeV}$ so that pions contribute less than $\sim 1 \%$ to the dark matter relic abundance and may be safely ignored. Then the total dark matter yield comes from photon annihilations alone and for $m_\chi \ll \TR$, the integrals in Eq.~\eqref{eq:Yinstan} may be done analytically, yielding:
\al{
Y_\text{DM} \approx \frac{3888 \sqrt{10}}{\pi^8} \frac{\Mpl}{g_{s,\ast} \sqrt{g_\ast}} \prn{\frac{17 y_n y_\chi \alpha}{4 \pi m_n m_\phi^2}}^2 \TR^5.
}
In this region of parameter space, the observed dark matter abundance thus predicts 
\al{
\sigma_{\chi n} &\approx \left(\frac{4 \pi}{17 \alpha}\right)^2 \frac{\pi^7}{3888 \sqrt{10}} \frac{g_{s,\ast} \sqrt{g_\ast}}{\Mpl}
\frac{\Omega_\text{DM} \rho_{crit}}{s_0} \frac{m_n^2 m_\chi}{\TR^5} \nonumber \\
&\approx 5.5 \times 10^{-44} \text{ cm}^2 \prn{\frac{g_{s,\ast} \sqrt{g_\ast}}{10.76^{3/2}}} \left(\frac{m_{\chi}}{1 \; \rm{MeV}}\right) \left(\frac{10 \; \rm{MeV}}{\TR} \right)^5 .
\label{eq:sigapprox}
}
Thus, lower reheating temperatures predict potentially detectable cross sections.

The full predictions, including the pion contributions, for the direct detection cross section as a function of $m_\chi$ are shown in Fig.~\ref{fig:sigmachin}. The colored contours correspond to the same reheating temperatures as in Fig.~\ref{fig:ynovrmphi2}, the gray regions correspond to the same bounds, and $y_\psi = y_\chi = 1$ again. The LHC constraint again comes from direct searches for $\psi$ (see Eq.~\eqref{eq:ynbnd}), and the dashed gray line assumes instead a possible future bound of $m_\psi \gtrsim 2 \text{ TeV}$ from the LHC. 

Note that the approximation of Eq.~\eqref{eq:sigapprox} holds as expected when $\TR \lesssim 12 \text{ MeV}$ and $m_\chi \ll \TR$. As $m_\chi$ increases,  $m_\chi \gtrsim \TR$, the photons able to produce the $\chi$s start becoming Boltzmann suppressed, requiring larger couplings and resulting in a larger predicted cross section.  Also note the increased spacing between contours for $\TR \gtrsim 14 \text{ MeV}$, when pion annihilations start to dominate. 

Much of the viable parameter space corresponding to $6 \text{ MeV} \lesssim \TR \lesssim 16 \text{ MeV}$ will be testable by a host of proposed direct detection experiments whose projected sensitivities are shown with dashed gray lines \cite{Knapen:2016cue,Kurinsky:2019pgb,Trickle:2019nya,Griffin:2019mvc,Campbell-Deem:2019hdx,Coskuner:2021qxo,Campbell-Deem:2022fqm}. We emphasize the ability of this cosmological history to predict cross sections which will be detectable in the future for dark matter masses below $\mathcal{O}(10 \text{ MeV})$. In fact, dark matter may be as light as $200 \text{ keV}$ and still be detectable by a $1 \text{ kg}\cdot\text{yr}$ exposure of CsI~\cite{Griffin:2019mvc}.

In Fig.~\ref{fig:sigmachin}, we have chosen $m_\phi \ge 1 \text{ GeV}$ for simplicity. As shown in Fig.~\ref{fig:ynovrmphi2}, the LHC bound stops dominating for $m_\phi \lesssim 1 \text{ GeV}$. For smaller $\phi$ masses, the LHC constraint below the BBN constraint in Fig.~\ref{fig:sigmachin} would be absent. This would allow dark matter masses as light as $\sim 80 \text{ keV}$ to be probed by future detectors if $T_{RH} \sim 6 $ MeV. We do not present this possibility in detail, however. For such light $\phi$ masses, the contours for $\TR \gtrsim 8 \text{ MeV}$ are sensitive to $m_\phi$ due to the pion resonance in Fig.~\ref{fig:ynovrmphi2}. 

\section{UV freeze-in with finite reheating}
\label{sec:FiniteRH}

In the previous section, the total dark matter yield was obtained assuming that reheating happened instantaneously at the temperature $\TR$. However, reheating from perturbative inflaton decay may not be immediate;\footnote{In fact, thermalization of the inflaton decay products may not be immediate, which can also in principle impact dark matter production~\cite{Harigaya:2014waa,Harigaya:2019tzu}, but we have checked that this negligibly impacts our yield calculation for sufficiently low inflaton masses.} then the yield in Eq.~\eqref{eq:Yinstan} can significantly underestimate the actual dark matter yield \cite{Bernal:2019mhf, Garcia:2017tuj}. Indeed $\Tmax$ may be much larger than $\TR$; assuming a matter dominated era of pre-heating $\Tmax \simeq 0.5 \, \left( m_\Phi / \Gamma_\Phi \right)^{1/4} \, \TR$ (see e.g. \cite{Garcia:2017tuj}). Since 10-MeV-scale reheating corresponds to $\Gamma_\Phi \sim \mathcal{O}\left(10^{-23} \right) \, \text{GeV}$, a simple single-field inflation model with instantaneous reheating from perturbative decay achieving $\Tmax \sim \TR$  would imply $m_\Phi \sim \Gamma_\Phi$ is ultra-light. Having $\Tmax > \TR$ may allow additional inflationary scenarios and also open new possibilities for baryogenesis. In this section, we redo the dark matter yield calculation in the case $\Tmax > \TR$ and detail the predictions for direct detection.

\subsection{Freeze-in calculations}
\label{sec:FI2}

While a significant amount of dark matter can be produced at temperatures above $\TR$, it is also diluted by the subsequent entropy generated from inflaton decays. This interplay between enhanced production and entropy dilution means that the yield of dark matter generated between $\Tmax$ to $\TR$ is sensitive to the temperature dependence of the freeze-in process(es)~\cite{Garcia:2017tuj}. We find the contribution to the dark matter yield from $\Tmax$ to $\TR$ is
\beq
Y^{\rm non-inst.}_\text{DM}= 0.4 \times \frac{y_n^2 y_\chi^2}{m_\phi^4 m_n^2} \frac{27 \sqrt{10} \Mpl}{32 \pi^8} \frac{g^{5/2}_\ast(\TR)}{g_{s,\ast}(\TR)} \TR^7  \sum_{j=\gamma,\pi^\pm} \kappa_j^2 \int_{\TR}^{\Tmax} \frac{dT}{T^{12}} \frac{I(m_j)}{g^3_\ast(T)}, 
\label{eq:Yfinite}
\eeq
where the $I(m_{j})$ is defined in Eq.~(\ref{eq:Idef}).
For the pion contribution, $\Tmax$ is capped at the temperature of the QCD phase transition, $156.5 \text{ MeV}$~\cite{HotQCD:2018pds}. Above this temperature, gluon annihilations to dark matter occur, but we find these contribute negligibly to the final dark matter relic abundance. The pre-factor of $0.4$ is a correction that comes from numerically solving the Boltzmann equations~\cite{Garcia:2017tuj}. Combining this yield with the one from temperatures below $\TR$, found in Eq.~\eqref{eq:Yinstan}, gives the total yield of dark matter in inflationary scenarios with arbitrary $\Tmax$.

\begin{figure}[t!]
    \begin{center}
        \includegraphics[width=0.95\textwidth]{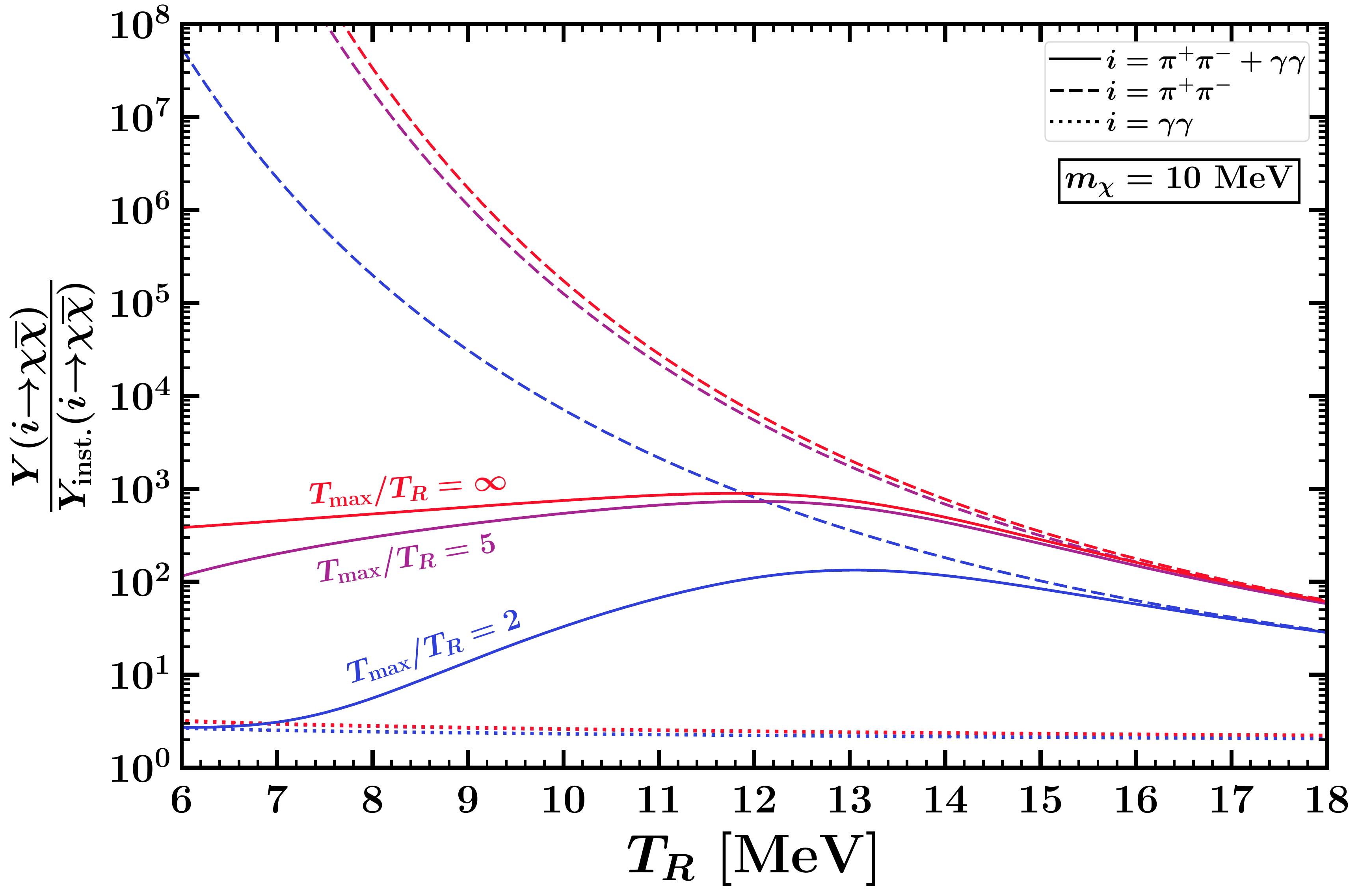}
    \end{center}
    \caption{The ratio of the total dark matter yield to the yield obtained only with the instantaneous reheating approximation, as a function of $\TR$ for $m_\chi = 10 \text{ MeV}$. The different colors correspond to different $\Tmax/\TR$, while the different line styles correspond to different sets of initial SM states.
    \label{fig:TotaltoInstantYield_vs_TR_varyTmaxbyTR}}
\end{figure}

For the small values of $\TR$ we consider, a larger $\Tmax$ implies exponentially more pions in the SM bath. The resulting temperature dependence of the thermally-averaged pion annihilation cross section can more than compensate for the entropy dilution between $\Tmax$ and $\TR$. Thus, the dark matter yield can be much larger than what we found assuming $\Tmax \simeq \TR$. 

To illustrate this point, in Fig.~\ref{fig:TotaltoInstantYield_vs_TR_varyTmaxbyTR} we show the ratio of the total dark matter yield, including the contribution from $\Tmax$ to $\TR$, to the yield obtained only with the instantaneous reheating approximation as a function of $\TR$. For concreteness, we have set $m_\chi = 10 \text{ MeV}$. The different colors correspond to different $\Tmax/\TR$, while the different line styles correspond to different sets of initial SM states. For $\Tmax/\TR = \infty$, we mean $\TR \ll \Tmax \lesssim \mathcal{O}(\text{TeV})$ since our calculation does not include the possible on-shell production of $\psi$'s. The solid lines show the full results obtained by including the contributions from both photon and pion annihilations. The dotted (dashed) lines show the results for photon (pion) contributions only. 

\begin{figure}[t!]
    \begin{center}
        \includegraphics[width=\textwidth]{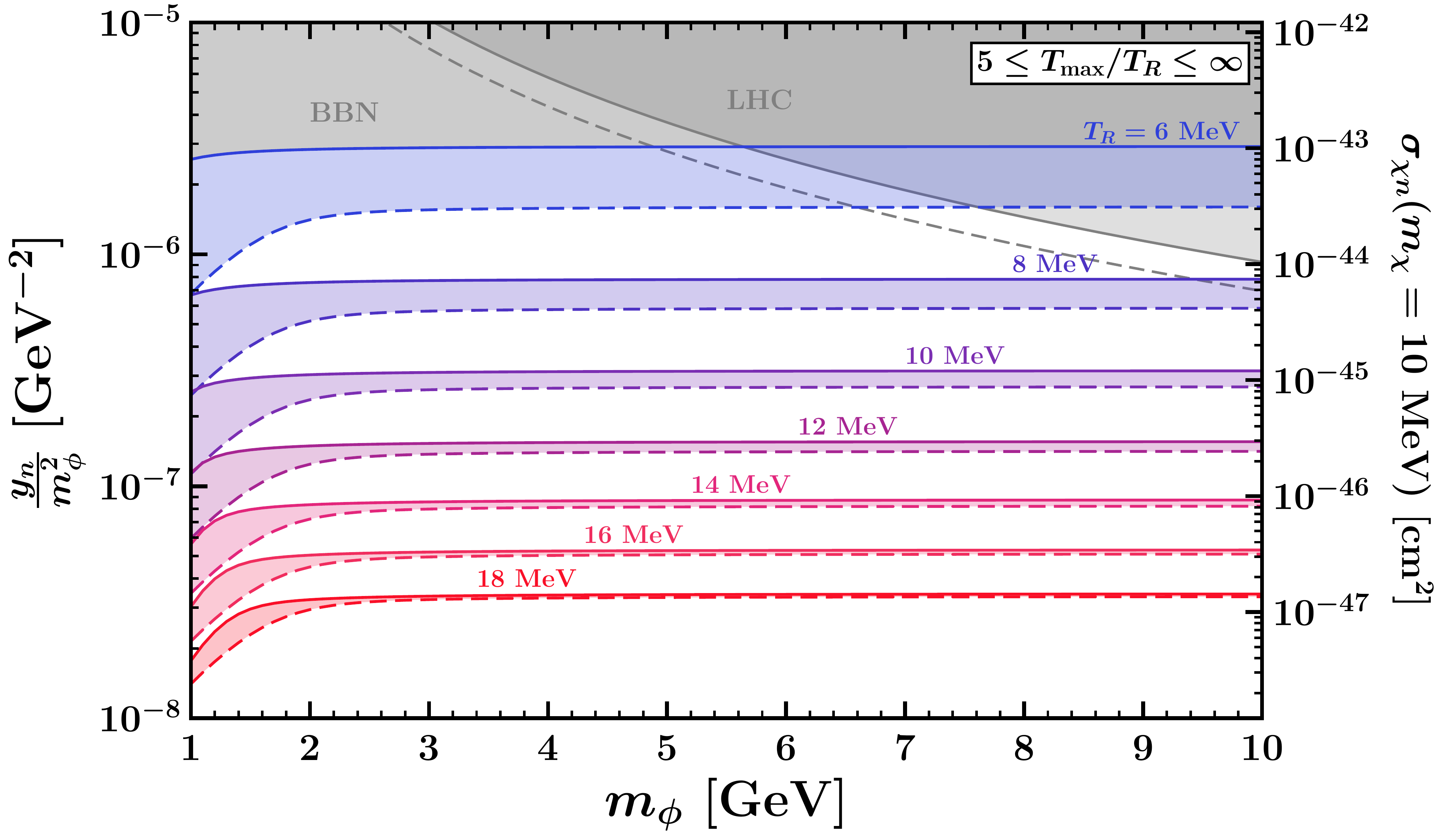}
    \end{center}
    \caption{
    Predictions for $y_n/m_{\phi}^2$ as a function of $m_\phi$ along with constraints in gray for $y_\psi = y_\chi = 1$. The colored bands yield the correct relic abundance for $m_\chi = 10 \text{ MeV}$ for various reheating temperatures (as labeled) and $5 \le \Tmax / \TR \le \infty$ (band width). The right vertical axis is the scattering cross section for $m_\chi = 10 \text{ MeV}$. 
    \label{fig:ynovrmphi2_Tmax}}
\end{figure}

The solid lines in Fig.~\ref{fig:TotaltoInstantYield_vs_TR_varyTmaxbyTR} prove that the total yield can be many orders of magnitude larger than expected from just the instantaneous reheating calculation, especially if $\Tmax$ is appreciably larger than $\TR$. As expected, this is due to the greatly enhanced contribution from pion annihilations, as is evident from the dashed contours. The dotted lines show that the contribution to the dark matter yield from photon annihilations from $\Tmax$ to $\TR$ is $\mathcal{O}(1)$ times its contribution from $\TR$ and below. Interestingly, the difference between the solid lines with $\Tmax/\TR = 5$ and $\infty$ is much less than between $\Tmax/\TR = 5$ and $2$. This, together with the previous observations, emphasizes that for large $\Tmax/\TR$, the dark matter relic abundance is set by pion annihilations above $\TR$. Furthermore, the contributions from the photon annihilations are always subdominant and can be ignored when $\Tmax \gtrsim 2 \TR$.

\subsection{Results}
\label{sec:results2}

\begin{figure}[t!]
    \begin{center}
        \includegraphics[width=0.95\textwidth]{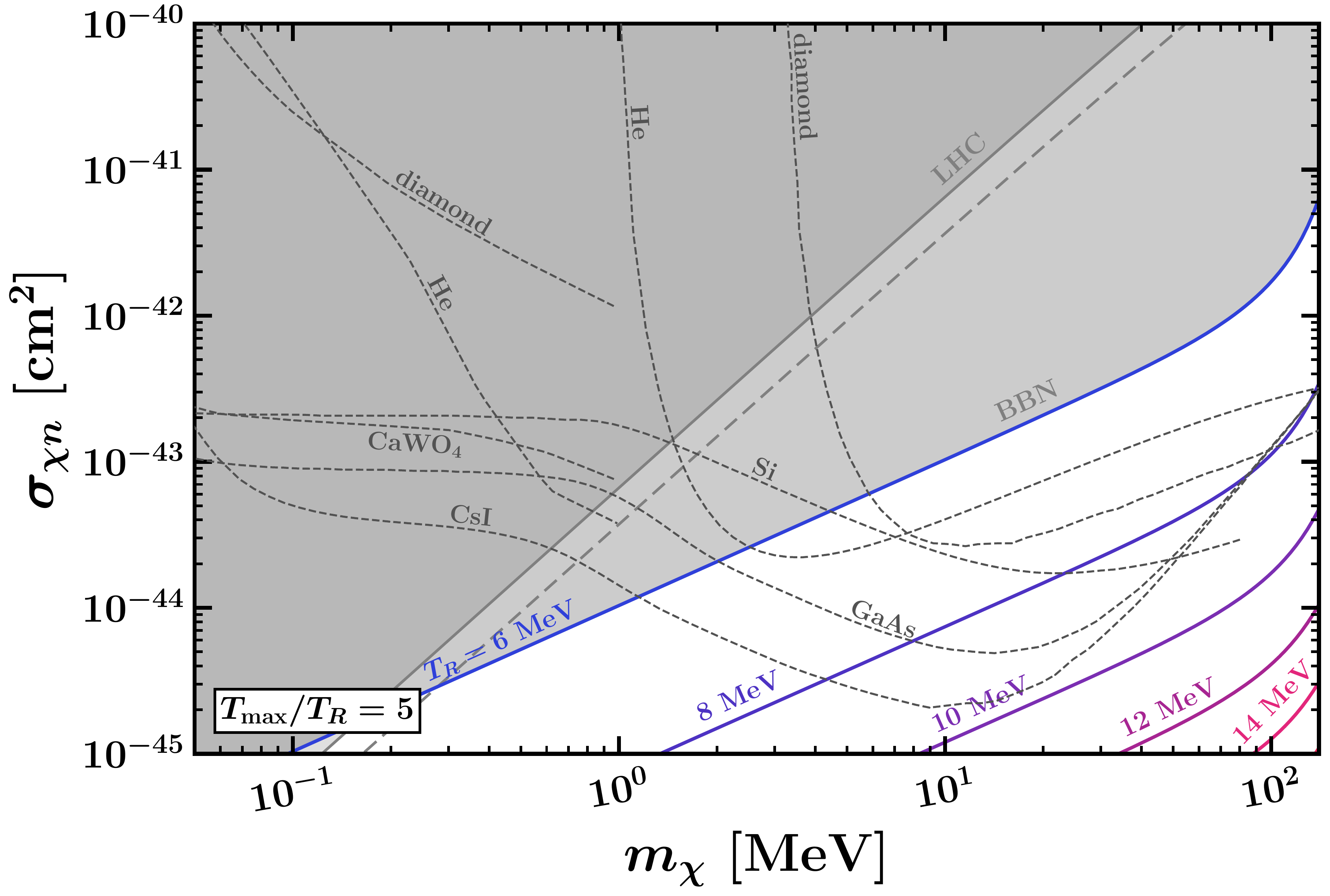}
    \end{center}
    \begin{center}
        \includegraphics[width=0.95\textwidth]{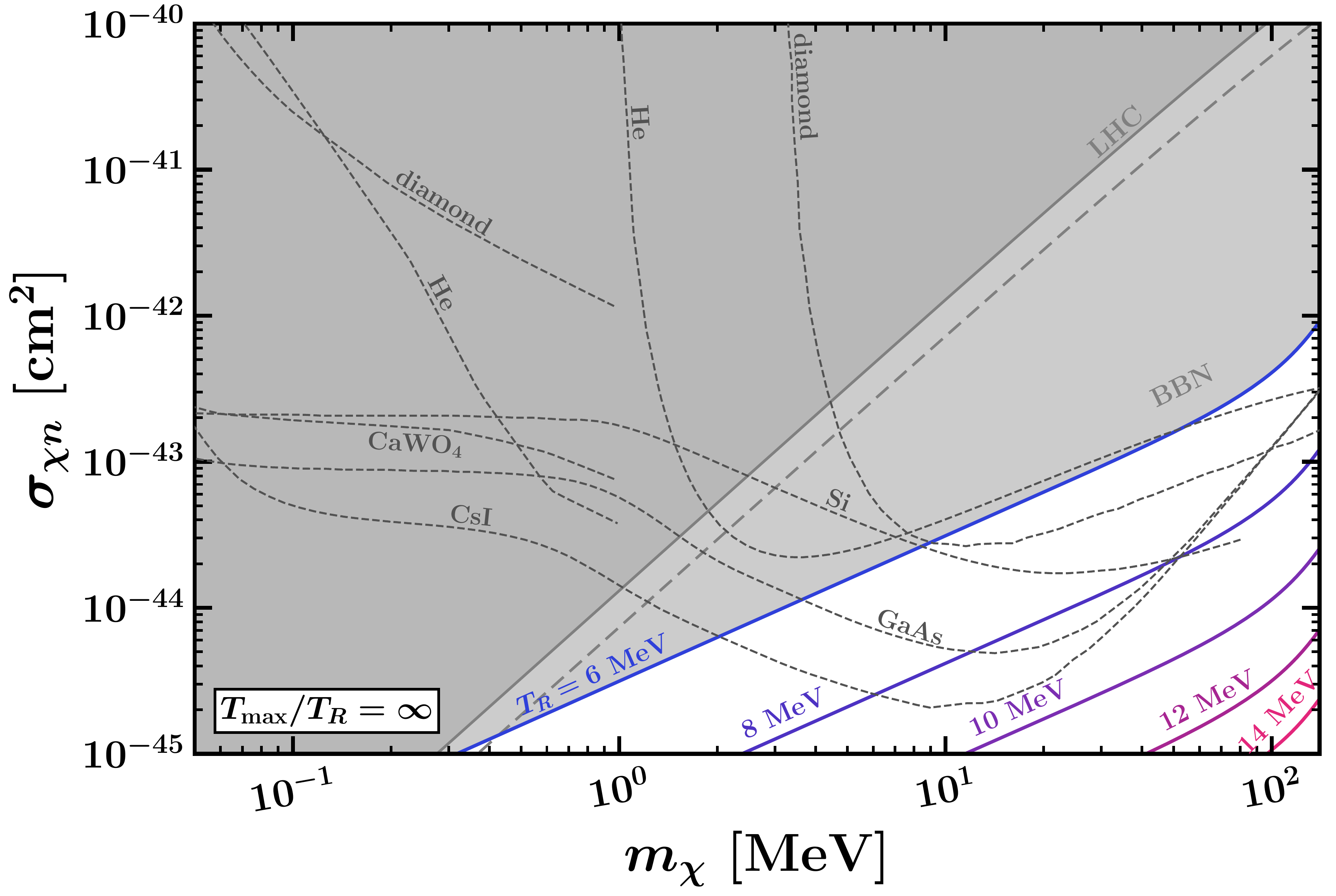}
    \end{center}
    \caption{
    The same as Fig.~\ref{fig:sigmachin}, but with $\Tmax/\TR = 5$ (top panel) or $\Tmax/\TR =\infty$ (bottom panel) instead of $\Tmax/\TR \simeq 1$.}
    \label{fig:sigmachin_Tmax}
\end{figure}

As in Sec.~\ref{sec:results1}, we plot the value of $y_n/m_\phi^2$ which gives the correct relic abundance as a function of $m_\phi$. In Fig.~\ref{fig:ynovrmphi2_Tmax}, this is done for a benchmark case of $(m_\chi = 10 \text{ MeV}, y_\psi=1, y_\chi=1)$ and various reheating temperatures. The bands for each $\TR$ value correspond to $5 \le \Tmax / \TR \le \infty$, with the top of each band $5$ and the bottom of each band $\infty$. Also shown in shades of gray are the same constraints from Fig.~\ref{fig:ynovrmphi2} and discussed in Sec.~\ref{sec:Model}. 

It is clear that the sensitivity to the pion resonance increases for larger $\Tmax/\TR$. We limit our parameter space scan by imposing $m_{\phi} \geq (2, 3)$ GeV for $\Tmax/\TR = (5, \infty)$, respectively, for the rest of this section. This both achieves larger cross sections and simplifies the analysis. As is evident from the figure, the requisite value of $y_n/m_\phi^2$ is then roughly independent of $m_\phi$. The right vertical axis of Fig.~\ref{fig:ynovrmphi2_Tmax} shows that direct detection cross sections, while still detectable for lower $\TR$, are greatly suppressed as $\Tmax/\TR$ increases, a result of the enhanced yield from pions. 

The full predictions, including the dominant contributions from $\Tmax$ to $\TR$, for the direct detection cross section as a function of $m_\chi$ are shown in Fig.~\ref{fig:sigmachin_Tmax} for both $\Tmax/\TR = 5$ (in the top panel) and $\Tmax/\TR =\infty$ (in the bottom panel). The colored contours, constraints, and future-experiment sensitivities correspond to those in Fig.~\ref{fig:sigmachin}. Again, we have set $y_\psi=y_\chi=1$. Notably, contours of fixed $\TR$ now occur at much lower values of $y_{n}/m_{\phi}^2$, and so the LHC constraint constrains a smaller area of the space allowed by BBN.

As expected, the effect of the enhanced freeze-in from $\Tmax$ to $\TR$ due to pion annihilations is to suppress the predicted direct detection cross section. The result is that future experiments will only be sensitive to a small range of parameter space for these inflationary histories. Still, there are benchmarks that predict detectable dark matter as light as $\sim 1 \text{ MeV}$ if $\Tmax / \TR =5$ or $\sim 2 \text{ MeV}$ if $\Tmax / \TR = \infty$.  Note also, for any inflationary scenario which results in a given $\TR$, the predicted cross section will be greater than or equal to the one predicted by $\Tmax / \TR = \infty$. 

\section{Discussion}
\label{sec:disc}

We have explored a model of hadrophilic, sub-GeV dark matter with cross sections detectable at proposed direct detection experiments. After taking into account all collider, BBN, and rare-meson-decay bounds, we find that UV freeze-in at $\mathcal{O}(10 \text{ MeV})$ reheating temperatures predicts sizable interactions for dark matter as light as $200 \text{ keV}$. This is significant because only a few dark matter models exist at such light masses which 1) may be detected at these proposed experiments and 2) are not already ruled out by cosmological or other constraints. Some proposals for detecting light dark matter have even resulted in experimental collaborations, e.g. using superfluid helium as a target~\cite{Hertel:2018aal,vonKrosigk:2022vnf}.  The existence of consistent benchmarks are important to support the science cases of these experiments.  

We have determined that even at such low reheating temperatures, contributions from pion annihilations can dominate the relic abundance. Thus, we have also shown that if the SM bath reaches a higher temperature prior to reheating, $\Tmax > \TR$, the corresponding cross sections are reduced. Even still, these more generic inflationary scenarios can predict benchmarks of relevance for direct detection proposals.

Given the generic challenges associated with constructing light dark matter models which are within the reach of the next generation of experiments, our low-reheating UV freeze-in scenario through a scalar mediator represents a significant benchmark for these proposals. For instance, two simple anomaly-free vector-mediator models, a kinetically mixed dark photon and gauged $B-L$, result in cross section many orders of magnitude below future sensitivities. This is due to the relative ease with which electron-positron pairs annihilate to produce dark matter through a dimension-six operator, in contrast with our scalar-mediator model.

In this work we found the enhanced UV freeze-in production from non-instantaneous inflaton decays to have a significant effect on the expected direct detection cross section. It would be interesting to consider dark matter production from UV freeze-in (and associated phenomenology) in more exotic models of reheating.\footnote{For reheating at high scales some work along these lines has been done for the case of Starobinsky Inflation \cite{Bernal:2020qyu} and NO models \cite{Bernal:2020bfj}.} 

Here, for simplicity, we have assumed that $\mathbf{Z}_2$ breaking is absent.  However, this need not be the case, and if present, would potentially allow Yukawa couplings between Standard Model fermions and the $\psi$ fields.  Such couplings could allow $\psi$ to impact precision electroweak observables, and could also change the details of the bounds that come, e.g. from meson decays, or direct LHC searches.  It would be interesting to consider the phenomenology of $\mathbf{Z}_2$-breaking scenarios in more detail.

\acknowledgments
We thank Valerie Domcke, Keisuke Harigaya, Stephen P.~Martin, Olcyr Sumensari, and  Jorinde van de Vis for useful discussions. R.M and A.P are supported in part by the DoE grant DE-SC0007859. G.E. is supported by the Cluster of Excellence {\em Precision Physics, Fundamental Interactions and Structure of Matter\/} (PRISMA${}^+$ -- EXC~2118/1) within the German Excellence Strategy (project ID 39083149). This research is also supported in part through computational resources and services provided by Advanced Research Computing (ARC), a division of Information and Technology Services (ITS) at the University of Michigan, Ann Arbor.

\appendix

\section{Alternative UV completions of hadrophilic dark matter}
\label{sec:altUV}

In this appendix, we consider alternative UV completions of the hadrophilic dark matter model. In particular, we consider having one or more vectorlike pairs of quarks and leptons, $\psi_i + \overline{\psi}_i$, in various representations under the SM gauge group, and we will study how the phenomenology depends upon the choice of representation. The generalized version of the Lagrangian in Eq.~(\ref{eq:Lagr}) includes 
\beq
\mathcal{L} \supset \sum_i - m_{\psi_i} \overline{\psi}_i \psi_i - y_{\psi_i} \phi \overline{\psi}_i \psi_i.
\label{eq:Lagr_altUV}
\eeq

The induced coupling of $\phi$ to gluons, after integrating out the colored $\psi_i + \overline{\psi}_i$, is now given by:
\beq
\mathcal{L} &\supset&
\frac{\alpha_s}{6 \pi}
\left (
\sum_{i} I_{c}(\psi_i) d_{L}(\psi_i) \frac{y_{\psi_i}}{m_{\psi_i}}
\right )
\phi G^a_{\mu \nu} G^{\mu \nu,a}
,
\label{eq:phiGG_phipsipsi}
\eeq
where $I_c({\psi_i})$ is the Dynkin index of the $\text{SU}(3)_c$ representation of $\psi_i$, $d_L({\psi_i})$ is the dimension of $\psi_i$ in the $\text{SU}(2)_L$ weak-isospin space. The coupling to gluons can then be mapped onto the low-energy nucleon coupling $y_n$ via\footnote{More generally, if $1/\Lambda_g$ is the coefficient of the $\phi G^a_{\mu \nu} G^{\mu \nu,a}$ operator, then the low-energy nucleon coupling is given by $y_n = \frac{8 \pi m_n}{b_0 \alpha_s \Lambda_g}$. Here $b_0 = 11 - \frac{2}{3} n_L$ is the coefficient of the QCD $\beta$-function at one-loop with $n_L$ being the number of quarks lighter than $\Lambda_{\rm QCD}$ and $m_\phi$ \cite{Gunion:1989we}.}:
\beq
y_n &=& \frac{4 m_n}{27} \sum_{i} I_c(\psi_i) d_L({\psi_i}) \frac{y_{\psi_i}}{m_{\psi_i}}.
\label{eq:phinn_phipsipsi}
\eeq

Electromagnetically charged $\psi_i$ also induce a one-loop coupling of $\phi$ to photons:
\beq
\mathcal{L} &\supset&
\frac{\alpha}{6 \pi}
\left (
\sum_i d_c({\psi_i}) {\rm Tr}[q^2({\psi_i})] \frac{y_{\psi_i}}{m_{\psi_i}}
\right )
\phi F_{\mu \nu} F^{\mu \nu}
,
\label{eq:phiFF_phipsipsi_ypsii}
\eeq
where $q({\psi_i})$ is the electromagnetic charge matrix of $\psi_i$ in the weak-isospin space, and $d_c({\psi_i})$ is the dimension of $\psi_i$ in $\text{SU}(3)_c$ color space. Due to the coupling to gluons in Eq.~(\ref{eq:phiGG_phipsipsi}), $\phi$ also has induced couplings to other charged hadrons, which together with the coupling to protons in Eq.~(\ref{eq:phinn_phipsipsi}), generate a contribution to the $\phi F_{\mu \nu} F^{\mu \nu}$ operator in the IR-- this is in addition to the contribution from integrating out $\psi_i$'s that carry an electromagnetic charge. Therefore, the coupling of $\phi$ to photons can be parameterized in terms of the nucleon coupling, as:
\beq
\mathcal{L}_{\phi F F} &=&
\left ( c_\gamma^{\rm IR} + c_\gamma^{\rm UV} \right )
\frac{y_n \alpha}{4 \pi m_n}
\phi F_{\mu \nu} F^{\mu \nu},
\label{eq:phiFF_phipsipsi}
\eeq
where
\beq
c_\gamma^{\rm IR} &=& 1 \quad \text{(NDA estimate)},
\eeq
accounts for the contribution from IR physics (charged hadrons), while $c_\gamma^{\rm UV}$ accounts for the contribution from UV physics (heavy vectorlike quarks). From eqs.~(\ref{eq:phinn_phipsipsi}) - (\ref{eq:phiFF_phipsipsi}), $c_\gamma^{\rm UV}$ can be computed in terms of $I_c$, $d_c$, $d_L$, ${\rm Tr}[q^2]$, and $y_{\psi_i}/m_{\psi_i}$ of various $\psi_i$.

The total dark matter yield obtained from pions and photons from temperatures below $\TR$ (from $\Tmax$ to $\TR$) can be computed using Eq.~(\ref{eq:Yinstan}) (Eq.~(\ref{eq:Yfinite})) with $\kappa_{\pi^\pm} = 1$ for the pions and $\kappa_{\gamma} = \frac{\alpha}{2 \pi} \left ( c_\gamma^{\rm IR} + c_\gamma^{\rm UV} \right )$, generalizing the contribution from photons in Eq.~(\ref{eq:kappa_yield_repQ}). As discussed in the main text, the photon contributions are negligible compared to the pion contributions for large $\TR$ ($\Tmax$). In the case where $\Tmax \simeq \TR$ (i.e. instantaneous reheating), for $\TR \lesssim 12$ MeV where the pion contributions can be ignored $\sigma_{\chi n}$ is $\left( \frac{17}{2} \right)^2 \frac{1}{(c_\gamma^{\rm IR} + c_\gamma^{\rm UV})^2}$ times the result given in Eq.~(\ref{eq:sigapprox}) for $m_\chi \ll \TR$.

\subsection*{UV completion with a single pair of $\psi + \overline{\psi}$:}
If we restrict ourselves to the case where there is only one vectorlike quark pair $\psi + \overline{\psi}$ that couples to $\phi$ as in Eq.~(\ref{eq:Lagr}), then the UV contribution to $\phi F_{\mu \nu} F^{\mu \nu}$ operator is given by
\beq
c_\gamma^{\rm UV} &=& 27 \frac{{\rm Tr}[q^2(\psi)]}{d_L(\psi)}
\:\qquad \text{($\phi \overline{\psi} \psi$; Eq.~(\ref{eq:Lagr}))},
\label{eq:cgamma_psi}
\eeq
independent of $y_\psi/m_\psi$. In the main text, $\psi$ is taken to transform as SM-like weak-isodoublet $Q \sim ({\bf 3}, {\bf 2}, \frac{1}{6})$ under the SM gauge group for which $c_\gamma^{\rm UV} = 15/2$. For other SM-like representations for $\psi$, namely up-type weak-isosinglet $U \sim ({\bf 3}, {\bf 1}, \frac{2}{3})$ and down-type weak-isosinglet $D \sim ({\bf 3}, {\bf 1}, -\frac{1}{3})$, $c_\gamma^{\rm UV} = 12$ and $3$, respectively.

Since, by assumption, the vectorlike quarks in these UV completions do not mix with the SM quarks, they can be long-lived and stable over detector lengths. Therefore, $m_\psi$ for various representations of $\psi$ can be constrained from the searches for pair-production of quasi-stable vectorlike quarks at particle colliders. In particular, we can infer bounds on up-type (down-type) vectorlike quarks from the LHC searches for long-lived top (bottom) squarks, by the ATLAS collaboration \cite{ATLAS:2019gqq}, to be
\beq
m_\psi &\gtrsim& 1.6 \text{ TeV} \quad \text{(95\% CL; up-type $\psi$)},\label{eq:Tbound}\\
m_\psi &\gtrsim& 1.5 \text{ TeV} \quad \text{(95\% CL; down-type $\psi$)},\label{eq:Bbound}
\eeq
based on ionization energy loss and time of flight.

As discussed in the main text, the strongest upper bound on the nucleon coupling $y_n$ comes from the lower bound on $m_\psi$. Here, the $y_n$ bound is given by
\begin{equation}
y_{n} = 9.3 \times 10^{-5} \left ( \frac{d_L(\psi)}{2} \right) \left(\frac{y_{\psi}}{1}\right)\left(\frac{1.5 \, \rm{TeV}}{m_{\psi}}\right).
\label{eq:ynbnd_varypsi}
\end{equation}
If reheating is assumed to be instantaneous (i.e. $\Tmax \simeq \TR$), it is evident from Fig.~\ref{fig:InPi2PhYvsTR} that the specific representation of $\psi$ somewhat matters only when $\TR$ is less than around 12 MeV, as the pion annihilations start dominating for larger $\TR$. In comparison with the results in Fig.~\ref{fig:sigmachin}, the colored contours for $\TR \lesssim 12$ MeV move slightly (down) up if $\psi$ transforms as ($U$) $D$ instead of $Q$. Specifically, for $m_\chi \ll \TR$, the cross sections $\sigma_{\chi n}$ that yield the correct relic abundance for UV completions with $U + \overline{U}$ and $D + \overline{D}$ are $(17/26)^2$ and $(17/8)^2$ times the corresponding cross sections for $Q + \overline{Q}$ in Fig.~\ref{fig:sigmachin}. The bounds on $\sigma_{\chi n}$ for up-type and down-type weak-isosinglets, based on Eq.~(\ref{eq:ynbnd_varypsi}), become stronger by a factor of 4 compared to the exclusion region labeled as LHC in Fig.~\ref{fig:sigmachin} for the SM-like weak-isodoublet quark representation.

Finally, we also comment on what happens if $\psi$ has more exotic quantum numbers: $\psi_0 \sim ({\bf 3}, {\bf 1}, 0)$. Since $\psi_0$ is an electroweak-singlet, $c_\gamma^{\rm UV} = 0$ from Eq.~(\ref{eq:cgamma_psi}), and therefore the $\phi F_{\mu \nu} F^{\mu \nu}$ operator is solely generated  from IR physics (i.e. charged hadrons). As a result, the dark matter yield from photon annihilations becomes more sensitive to the uncertainty in the NDA estimate of $c_\gamma^{\rm IR} = 1$. 

In this case, hadronization will result in fractionally charged hadrons which are not presently the target of a dedicated search at ATLAS/CMS. And while limits relying on $dE/dx$ do not immediately apply, searches looking for slow moving particles using time of flight measurements should. Based on the ATLAS searches for long-lived $R$-hadrons using the time of flight measurements only \cite{ATLAS:2019gqq}, we estimate: 
\beq
m_{\psi} & \gtrsim & 1.5 \text{ to } 1.7 \text{ TeV} \quad \text{(95\% CL; electroweak-singlet $\psi$)},
\label{eq:Psi0bound}
\eeq
where the range of values comes from assuming a signal selection efficiency that is a factor of 2 higher or lower than that of the long-lived top-squark searches. Here, we obtained the observed signal upper limits using the {\sc Zstats} package \cite{Zstats} based on Bayesian-motivated statistical measures in \cite{Bhattiprolu:2022xhm}. A dedicated search might extend these limits slightly. Note that $\sigma_{\chi n}$ with $\psi \sim \psi_0$, for $m_\chi \ll \TR$ and $\TR \lesssim 12$ MeV, is enhanced by a factor of $(17/2)^2$ compared to the corresponding contours in Fig.~\ref{fig:sigmachin}. The bound on $\sigma_{\chi n}$ using the $y_n$ bound of Eq.~(\ref{eq:ynbnd_varypsi}) and $m_\psi \gtrsim 1.7$ TeV is roughly a factor of $4$ stronger than the corresponding exclusion labeled as LHC in Fig.~\ref{fig:sigmachin}.

\subsection*{UV completion with a  full GUT multiplet of vectorlike fermions:}
Finally, we consider UV completions with vectorlike fermions in ${\bf 5 + \overline{5}}$ and ${\bf 10 + \overline{10}}$ representations of $\text{SU}(5)$, where ${\bf \overline{5}} = \overline{D} + L$ and ${\bf 10} = Q + \overline{U} + \overline{E}$ contain all of SM quark and lepton representations. Here, $L + \overline{L}$ and $E + \overline{E}$ are $\text{SU}(2)_L$ doublet and (charged) singlet vectorlike leptons, respectively. This possibility might be motivated for possible embedding in a GUT framework, and could preserve the apparent gauge coupling unification if we assume supersymmetry. We, however, do not assume an unbroken $\text{SU}(5)$ gauge symmetry in the UV.

In the special case where $y_{\psi_i}/m_{\psi_i}$ are the same for all vectorlike fermions, we obtain the UV contribution to $\phi F_{\mu \nu} F^{\mu \nu}$ operator at low energy to be:
\beq
c_\gamma^{\rm UV} &=& \frac{9}{2} \dfrac{\sum\limits_i d_c({\psi_i}) {\rm Tr}[q^2({\psi_i})]}{\sum\limits_{i} I_c(\psi_i) d_L({\psi_i})}
\:\qquad \text{($\phi \overline{\psi}_i \psi_i$ with same $\frac{y_{\psi_i}}{m_{\psi_i}}$; Eq.~(\ref{eq:Lagr_altUV}))}.
\eeq
For vectorlike fermions in ${\bf 5 + \overline{5}}$ and/or ${ \bf10 + \overline{10}}$ of $\text{SU}(5)$ with the same $y_{\psi_i}/m_{\psi_i}$, $c_\gamma^{\rm UV} = 12$. In a realistic GUT scenario, however, $y_{\psi_i}/m_{\psi_i}$ for various vectorlike fermions in a GUT multiplet at low energies, after the renormalization group evolution from the GUT scale, would not be same. After specifying a GUT, it is  straightforward to obtain $c_\gamma^{\rm UV}$ in terms of $I_c$, $d_c$, $d_L$, ${\rm Tr}[q^2]$, and $y_{\psi_i}/m_{\psi_i}$ of various $\psi_i$.

\begin{figure}[H]
    \begin{center}
        \includegraphics[width=0.95\textwidth]{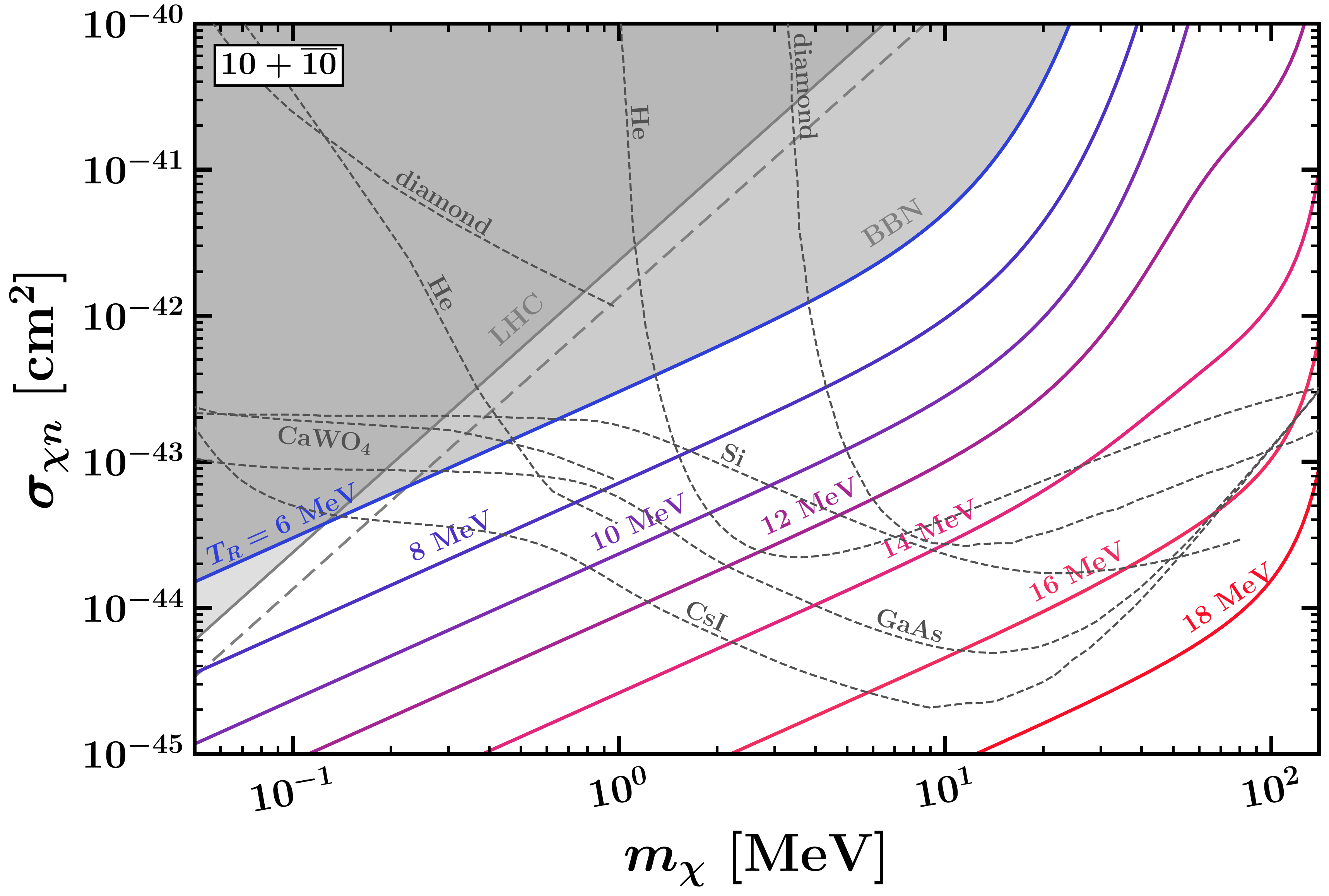}
    \end{center}
    \begin{center}
        \includegraphics[width=0.95\textwidth]{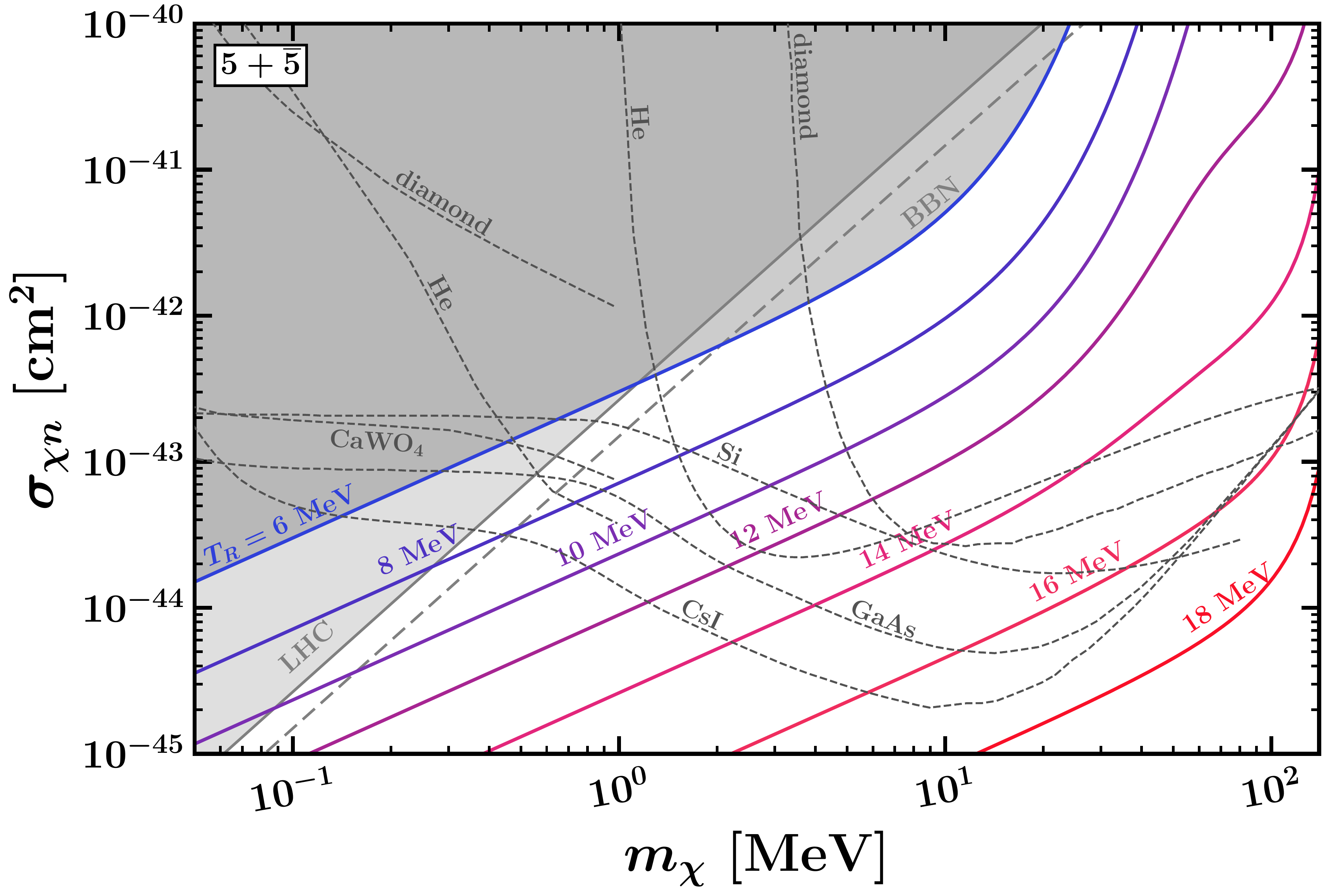}
    \end{center}
    \caption{The dark matter-nucleon cross section $\sigma_{\chi n}$ as a function of dark matter mass $m_\chi$ for $y_{\psi_i} = y_\chi = 1$. For reheating temperatures shown, each colored contour yields the correct dark matter abundance for heavy TeV-scale vectorlike fermions in the $5 + \overline{5}$ (top panel) and $10 + \overline{10}$ (bottom panel) representations of $\text{SU}(5)$,  see text for additional simplifying assumptions. The gray shaded regions are excluded by the LHC and BBN, while the dashed gray lines show the projected sensitivities of future experiments \cite{Knapen:2016cue,Kurinsky:2019pgb,Trickle:2019nya,Griffin:2019mvc,Campbell-Deem:2019hdx,Coskuner:2021qxo,Campbell-Deem:2022fqm}.}
    \label{fig:sigmachin_SU5reps}
\end{figure}

Fig.~\ref{fig:sigmachin_SU5reps} shows the predicted direct detection cross section $\sigma_{\chi n}$ as a function of the dark matter mass $m_\chi$ for heavy vectorlike fermions in complete $\text{SU}(5)$ representations: ${ \bf10 + \overline{10}}$ in the top panel and ${\bf 5 + \overline{5}}$ in the bottom panel. For simplicity and illustration purposes, absent a concrete GUT model, we chose $y_{\psi_i}/m_{\psi_i} (\equiv y_{\text{SU}(5)}/m_{\text{SU}(5)})$ to be the equal at the IR scale for all vectorlike content in a chosen $\text{SU}(5)$ pair and set $y_{\psi_i} = y_\chi = 1$. The colored contours correspond to various reheating temperatures from 6 to 18 MeV. The constraints and future experimental sensitivities labeled in the plot correspond to the ones shown in Fig.~\ref{fig:sigmachin}. The constraint labeled as LHC, in particular, is obtained from an analog of Equation (\ref{eq:ynbnd_varypsi}), with $d_L({\psi})/2$ replaced by $\sum_{i} I_c(\psi_i) d_L({\psi_i})$, with $y_{\text{SU}(5)} \lesssim 1$ constrained only by perturbativity and $m_{\text{SU}(5)} \gtrsim 1.5$ ($2$) TeV for shaded region (long dashed line).
Under these assumptions, we can note the slight increase (decrease) in the parameter space
for ${\bf 10 + \overline{10}}$ (${ \bf5 + \overline{5}}$) of $\text{SU}(5)$ compared to Fig.~\ref{fig:sigmachin}.

\newpage

\section{Finite Reheating Yield Calculation}
\label{sec:FiniteReheatingYield}

This Appendix derives the yield of dark matter due to a period of finite reheating found in Eq.~\eqref{eq:Yfinite}. We make use of the notation and derivations in \cite{Garcia:2017tuj}. There are two major changes in this non-instantaneous reheating derivation relative to the instantaneous one: 1) the usual (approximate) $\partial T / \partial t = - HT$ no longer holds and 2) $Y_\chi = n_\chi / s$ is no longer a useful parameterization since bath entropy is not conserved as the inflaton decays. Updating the first relation, we find that while the inflaton is decaying:
\al{
\frac{\partial T}{\partial t} = -\frac{T \Gamma_\Phi}{4v}, \text{where } v=\Gamma_\Phi \prn{t-t_\text{end}}.
\label{eq:Tdotnoninst}
}
$\Gamma_\Phi$ is the inflaton decay rate and the ``end'' subscript corresponds to the end of inflation. We have in mind a single-field inflationary scenario and assume that at the end of inflation, the inflaton undergoes coherent oscillations about its potential minimum. 

As for the second change, it is useful to note that:
\al{
\frac{d}{dT} \prn{n_\chi \prn{\frac{a}{a_\text{end}}}^3}=\prn{\dot{n}_\chi + 3 H n_\chi} \frac{1}{\dot{T}} \prn{\frac{a}{a_\text{end}}}^3.
}
Using Eq.~\eqref{eq:Tdotnoninst}, noting the number-density Boltzmann equation for $\chi$, and integrating over temperatures, we find
\al{
n_\chi (\TR) \prn{\frac{a(\TR)}{a_\text{end}}}^3 \! = \! \int_{\TR}^{\Tmax} \!\!\!\!\!\! dT \left[\avg{\sigma_{\chi \overline{\chi} \to \gamma \gamma} v} + \avg{\sigma_{\chi \overline{\chi} \to \pi^+ \pi^-} v} \right] \prn{n_\chi^\text{eq}}^2 \frac{4 v}{T \Gamma_\Phi} \prn{\frac{a}{a_\text{end}}}^3,
\label{eq:nxnoninst}
}
where the left hand side is evaluated at the reheating temperature. Assuming that the inflaton dominates the energy density of the Universe until the end of reheating allows us to rewrite
\al{
\prn{\frac{a}{a_\text{end}}}^3 \simeq \prn{\frac{v}{A}}^2, \text{where } A \simeq \mathcal{O}(1) \frac{\Gamma_\Phi}{m_\Phi}.
\label{eq:aovraend}
}
The $\mathcal{O}(1)$ number depends on the particular inflationary model, but drops out from the final yield expression. We further assume that $A \ll v \ll 1$ to find that
\al{
v \simeq \frac{24 \Gamma_\Phi^2 \Mpl^2}{\pi^2 g_\ast T^4}.
\label{eq:vappr}
}
Plugging Eqs.~\eqref{eq:aovraend} and \eqref{eq:vappr} into Eq.~\eqref{eq:nxnoninst} and dividing by the bath entropy at the reheating temperature immediately gives Eq.~\eqref{eq:Yfinite}.

\bibliographystyle{JHEP}
\bibliography{ref}

\providecommand{\href}[2]{#2}\begingroup\raggedright\begin{thebibliography}{10}

\bibitem{XENON:2020kmp}
{\bf XENON} Collaboration, E.~Aprile et~al., {\it {Projected WIMP sensitivity
  of the XENONnT dark matter experiment}},  {\em JCAP} {\bf 11} (2020) 031,
  [\href{http://arxiv.org/abs/2007.08796}{{\tt arXiv:2007.08796}}].

\bibitem{Aalbers:2022fxq}
J.~Aalbers et~al., {\it {First Dark Matter Search Results from the LUX-ZEPLIN
  (LZ) Experiment}},  \href{http://arxiv.org/abs/2207.03764}{{\tt
  arXiv:2207.03764}}.

\bibitem{PandaX-4T:2021bab}
{\bf PandaX-4T} Collaboration, Y.~Meng et~al., {\it {Dark Matter Search Results
  from the PandaX-4T Commissioning Run}},  {\em Phys. Rev. Lett.} {\bf 127}
  (2021), no.~26 261802, [\href{http://arxiv.org/abs/2107.13438}{{\tt
  arXiv:2107.13438}}].

\bibitem{Ballesteros:2020adh}
G.~Ballesteros, M.~A.~G. Garcia, and M.~Pierre, {\it {How warm are non-thermal
  relics? Lyman-$\alpha$ bounds on out-of-equilibrium dark matter}},  {\em
  JCAP} {\bf 03} (2021) 101, [\href{http://arxiv.org/abs/2011.13458}{{\tt
  arXiv:2011.13458}}].

\bibitem{Knapen:2016cue}
S.~Knapen, T.~Lin, and K.~M. Zurek, {\it {Light Dark Matter in Superfluid
  Helium: Detection with Multi-excitation Production}},  {\em Phys. Rev. D}
  {\bf 95} (2017), no.~5 056019, [\href{http://arxiv.org/abs/1611.06228}{{\tt
  arXiv:1611.06228}}].

\bibitem{Budnik:2017sbu}
R.~Budnik, O.~Chesnovsky, O.~Slone, and T.~Volansky, {\it {Direct Detection of
  Light Dark Matter and Solar Neutrinos via Color Center Production in
  Crystals}},  {\em Phys. Lett. B} {\bf 782} (2018) 242--250,
  [\href{http://arxiv.org/abs/1705.03016}{{\tt arXiv:1705.03016}}].

\bibitem{Knapen:2017xzo}
S.~Knapen, T.~Lin, and K.~M. Zurek, {\it {Light Dark Matter: Models and
  Constraints}},  {\em Phys. Rev. D} {\bf 96} (2017), no.~11 115021,
  [\href{http://arxiv.org/abs/1709.07882}{{\tt arXiv:1709.07882}}].

\bibitem{Knapen:2017ekk}
S.~Knapen, T.~Lin, M.~Pyle, and K.~M. Zurek, {\it {Detection of Light Dark
  Matter With Optical Phonons in Polar Materials}},  {\em Phys. Lett. B} {\bf
  785} (2018) 386--390, [\href{http://arxiv.org/abs/1712.06598}{{\tt
  arXiv:1712.06598}}].

\bibitem{Griffin:2018bjn}
S.~Griffin, S.~Knapen, T.~Lin, and K.~M. Zurek, {\it {Directional Detection of
  Light Dark Matter with Polar Materials}},  {\em Phys. Rev. D} {\bf 98}
  (2018), no.~11 115034, [\href{http://arxiv.org/abs/1807.10291}{{\tt
  arXiv:1807.10291}}].

\bibitem{Kurinsky:2019pgb}
N.~A. Kurinsky, T.~C. Yu, Y.~Hochberg, and B.~Cabrera, {\it {Diamond Detectors
  for Direct Detection of Sub-GeV Dark Matter}},  {\em Phys. Rev. D} {\bf 99}
  (2019), no.~12 123005, [\href{http://arxiv.org/abs/1901.07569}{{\tt
  arXiv:1901.07569}}].

\bibitem{Essig:2019kfe}
R.~Essig, J.~P\'erez-R\'\i{}os, H.~Ramani, and O.~Slone, {\it {Direct Detection
  of Spin-(In)dependent Nuclear Scattering of Sub-GeV Dark Matter Using
  Molecular Excitations}},  {\em Phys. Rev. Research.} {\bf 1} (2019) 033105,
  [\href{http://arxiv.org/abs/1907.07682}{{\tt arXiv:1907.07682}}].

\bibitem{Trickle:2019nya}
T.~Trickle, Z.~Zhang, K.~M. Zurek, K.~Inzani, and S.~Griffin, {\it
  {Multi-Channel Direct Detection of Light Dark Matter: Theoretical
  Framework}},  {\em JHEP} {\bf 03} (2020) 036,
  [\href{http://arxiv.org/abs/1910.08092}{{\tt arXiv:1910.08092}}].

\bibitem{Griffin:2019mvc}
S.~M. Griffin, K.~Inzani, T.~Trickle, Z.~Zhang, and K.~M. Zurek, {\it
  {Multichannel direct detection of light dark matter: Target comparison}},
  {\em Phys. Rev. D} {\bf 101} (2020), no.~5 055004,
  [\href{http://arxiv.org/abs/1910.10716}{{\tt arXiv:1910.10716}}].

\bibitem{Campbell-Deem:2019hdx}
B.~Campbell-Deem, P.~Cox, S.~Knapen, T.~Lin, and T.~Melia, {\it {Multiphonon
  excitations from dark matter scattering in crystals}},  {\em Phys. Rev. D}
  {\bf 101} (2020), no.~3 036006, [\href{http://arxiv.org/abs/1911.03482}{{\tt
  arXiv:1911.03482}}]. [Erratum: Phys.Rev.D 102, 019904 (2020)].

\bibitem{Griffin:2020lgd}
S.~M. Griffin, Y.~Hochberg, K.~Inzani, N.~Kurinsky, T.~Lin, and T.~Chin, {\it
  {Silicon carbide detectors for sub-GeV dark matter}},  {\em Phys. Rev. D}
  {\bf 103} (2021), no.~7 075002, [\href{http://arxiv.org/abs/2008.08560}{{\tt
  arXiv:2008.08560}}].

\bibitem{Coskuner:2021qxo}
A.~Coskuner, T.~Trickle, Z.~Zhang, and K.~M. Zurek, {\it {Directional
  Detectability of Dark Matter With Single Phonon Excitations: Target
  Comparison}},  \href{http://arxiv.org/abs/2102.09567}{{\tt
  arXiv:2102.09567}}.

\bibitem{Campbell-Deem:2022fqm}
B.~Campbell-Deem, S.~Knapen, T.~Lin, and E.~Villarama, {\it {Dark matter direct
  detection from the single phonon to the nuclear recoil regime}},  {\em Phys.
  Rev. D} {\bf 106} (2022), no.~3 036019,
  [\href{http://arxiv.org/abs/2205.02250}{{\tt arXiv:2205.02250}}].

\bibitem{Coskuner:2018are}
A.~Coskuner, D.~M. Grabowska, S.~Knapen, and K.~M. Zurek, {\it {Direct
  Detection of Bound States of Asymmetric Dark Matter}},  {\em Phys. Rev. D}
  {\bf 100} (2019), no.~3 035025, [\href{http://arxiv.org/abs/1812.07573}{{\tt
  arXiv:1812.07573}}].

\bibitem{Smirnov:2020zwf}
J.~Smirnov and J.~F. Beacom, {\it {New Freezeout Mechanism for Strongly
  Interacting Dark Matter}},  {\em Phys. Rev. Lett.} {\bf 125} (2020), no.~13
  131301, [\href{http://arxiv.org/abs/2002.04038}{{\tt arXiv:2002.04038}}].

\bibitem{Elor:2021swj}
G.~Elor, R.~McGehee, and A.~Pierce, {\it {Maximizing Direct Detection with
  HYPER Dark Matter}},  \href{http://arxiv.org/abs/2112.03920}{{\tt
  arXiv:2112.03920}}.

\bibitem{An:2022sva}
R.~An, V.~Gluscevic, E.~Calabrese, and J.~C. Hill, {\it {What does cosmology
  tell us about the mass of thermal-relic dark matter?}},  {\em JCAP} {\bf 07}
  (2022), no.~07 002, [\href{http://arxiv.org/abs/2202.03515}{{\tt
  arXiv:2202.03515}}].

\bibitem{Hall:2009bx}
L.~J. Hall, K.~Jedamzik, J.~March-Russell, and S.~M. West, {\it {Freeze-In
  Production of FIMP Dark Matter}},  {\em JHEP} {\bf 03} (2010) 080,
  [\href{http://arxiv.org/abs/0911.1120}{{\tt arXiv:0911.1120}}].

\bibitem{Essig:2011nj}
R.~Essig, J.~Mardon, and T.~Volansky, {\it {Direct Detection of Sub-GeV Dark
  Matter}},  {\em Phys. Rev. D} {\bf 85} (2012) 076007,
  [\href{http://arxiv.org/abs/1108.5383}{{\tt arXiv:1108.5383}}].

\bibitem{SENSEI:2020dpa}
{\bf SENSEI} Collaboration, L.~Barak et~al., {\it {SENSEI: Direct-Detection
  Results on sub-GeV Dark Matter from a New Skipper-CCD}},  {\em Phys. Rev.
  Lett.} {\bf 125} (2020), no.~17 171802,
  [\href{http://arxiv.org/abs/2004.11378}{{\tt arXiv:2004.11378}}].

\bibitem{Elahi:2014fsa}
F.~Elahi, C.~Kolda, and J.~Unwin, {\it {UltraViolet Freeze-in}},  {\em JHEP}
  {\bf 03} (2015) 048, [\href{http://arxiv.org/abs/1410.6157}{{\tt
  arXiv:1410.6157}}].

\bibitem{Berlin:2018bsc}
A.~Berlin, N.~Blinov, G.~Krnjaic, P.~Schuster, and N.~Toro, {\it {Dark Matter,
  Millicharges, Axion and Scalar Particles, Gauge Bosons, and Other New Physics
  with LDMX}},  {\em Phys. Rev. D} {\bf 99} (2019), no.~7 075001,
  [\href{http://arxiv.org/abs/1807.01730}{{\tt arXiv:1807.01730}}].

\bibitem{Dimopoulos:1987rk}
S.~Dimopoulos and L.~J. Hall, {\it {Baryogenesis at the {MeV} Era}},  {\em
  Phys. Lett. B} {\bf 196} (1987) 135--141.

\bibitem{Cline:1990bw}
J.~M. Cline and S.~Raby, {\it {Gravitino induced baryogenesis: A Problem made a
  virtue}},  {\em Phys. Rev. D} {\bf 43} (1991) 1781--1787.

\bibitem{Aitken:2017wie}
K.~Aitken, D.~McKeen, T.~Neder, and A.~E. Nelson, {\it {Baryogenesis from
  Oscillations of Charmed or Beautiful Baryons}},  {\em Phys. Rev. D} {\bf 96}
  (2017), no.~7 075009, [\href{http://arxiv.org/abs/1708.01259}{{\tt
  arXiv:1708.01259}}].

\bibitem{Elor:2018twp}
G.~Elor, M.~Escudero, and A.~Nelson, {\it {Baryogenesis and Dark Matter from
  $B$ Mesons}},  {\em Phys. Rev. D} {\bf 99} (2019), no.~3 035031,
  [\href{http://arxiv.org/abs/1810.00880}{{\tt arXiv:1810.00880}}].

\bibitem{Elor:2020tkc}
G.~Elor and R.~McGehee, {\it {Making the Universe at 20 MeV}},  {\em Phys. Rev.
  D} {\bf 103} (2021), no.~3 035005,
  [\href{http://arxiv.org/abs/2011.06115}{{\tt arXiv:2011.06115}}].

\bibitem{Elahi:2021jia}
F.~Elahi, G.~Elor, and R.~McGehee, {\it {Charged B mesogenesis}},  {\em Phys.
  Rev. D} {\bf 105} (2022), no.~5 055024,
  [\href{http://arxiv.org/abs/2109.09751}{{\tt arXiv:2109.09751}}].

\bibitem{Jaeckel:2022osh}
J.~Jaeckel and W.~Yin, {\it {High Energy Sphalerons for Baryogenesis at Low
  Temperatures}},  \href{http://arxiv.org/abs/2206.06376}{{\tt
  arXiv:2206.06376}}.

\bibitem{ATLAS:2019gqq}
{\bf ATLAS} Collaboration, M.~Aaboud et~al., {\it {Search for heavy charged
  long-lived particles in the ATLAS detector in 36.1 fb$^{-1}$ of proton-proton
  collision data at $\sqrt{s} = 13$ TeV}},  {\em Phys. Rev. D} {\bf 99} (2019),
  no.~9 092007, [\href{http://arxiv.org/abs/1902.01636}{{\tt
  arXiv:1902.01636}}].

\bibitem{Gunion:1989we}
J.~F. Gunion, H.~E. Haber, G.~L. Kane, and S.~Dawson, {\em {The Higgs Hunter's
  Guide}}, vol.~80.
\newblock 2000.

\bibitem{Georgi:1986kr}
H.~Georgi and L.~Randall, {\it {Flavor Conserving CP Violation in Invisible
  Axion Models}},  {\em Nucl. Phys. B} {\bf 276} (1986) 241--252.

\bibitem{Manohar:1983md}
A.~Manohar and H.~Georgi, {\it {Chiral Quarks and the Nonrelativistic Quark
  Model}},  {\em Nucl. Phys. B} {\bf 234} (1984) 189--212.

\bibitem{Randall:2008ppe}
S.~W. Randall, M.~Markevitch, D.~Clowe, A.~H. Gonzalez, and M.~Bradac, {\it
  {Constraints on the Self-Interaction Cross-Section of Dark Matter from
  Numerical Simulations of the Merging Galaxy Cluster 1E 0657-56}},  {\em
  Astrophys. J.} {\bf 679} (2008) 1173--1180,
  [\href{http://arxiv.org/abs/0704.0261}{{\tt arXiv:0704.0261}}].

\bibitem{Kaplinghat:2015aga}
M.~Kaplinghat, S.~Tulin, and H.-B. Yu, {\it {Dark Matter Halos as Particle
  Colliders: Unified Solution to Small-Scale Structure Puzzles from Dwarfs to
  Clusters}},  {\em Phys. Rev. Lett.} {\bf 116} (2016), no.~4 041302,
  [\href{http://arxiv.org/abs/1508.03339}{{\tt arXiv:1508.03339}}].

\bibitem{BaBar:2013npw}
{\bf BaBar} Collaboration, J.~P. Lees et~al., {\it {Search for $B \to K^{(*)}
  \nu \overline \nu$ and invisible quarkonium decays}},  {\em Phys. Rev. D}
  {\bf 87} (2013), no.~11 112005, [\href{http://arxiv.org/abs/1303.7465}{{\tt
  arXiv:1303.7465}}].

\bibitem{NA62:2021zjw}
{\bf NA62} Collaboration, E.~Cortina~Gil et~al., {\it {Measurement of the very
  rare K$^{+}$\textrightarrow{}$ {\pi}^{+}\nu \overline{\nu} $ decay}},  {\em
  JHEP} {\bf 06} (2021) 093, [\href{http://arxiv.org/abs/2103.15389}{{\tt
  arXiv:2103.15389}}].

\bibitem{Charles:2020dfl}
J.~Charles, S.~Descotes-Genon, Z.~Ligeti, S.~Monteil, M.~Papucci, K.~Trabelsi,
  and L.~Vale~Silva, {\it {New physics in $B$ meson mixing: future sensitivity
  and limitations}},  {\em Phys. Rev. D} {\bf 102} (2020), no.~5 056023,
  [\href{http://arxiv.org/abs/2006.04824}{{\tt arXiv:2006.04824}}].

\bibitem{Kofman:1997pt}
L.~Kofman, {\it {Reheating and preheating after inflation}},  in {\em {3rd
  RESCEU International Symposium on Particle Cosmology}}, pp.~1--8, 11, 1997.
\newblock \href{http://arxiv.org/abs/hep-ph/9802285}{{\tt hep-ph/9802285}}.

\bibitem{Hannestad:2004px}
S.~Hannestad, {\it {What is the lowest possible reheating temperature?}},  {\em
  Phys. Rev. D} {\bf 70} (2004) 043506,
  [\href{http://arxiv.org/abs/astro-ph/0403291}{{\tt astro-ph/0403291}}].

\bibitem{deSalas:2015glj}
P.~F. de~Salas, M.~Lattanzi, G.~Mangano, G.~Miele, S.~Pastor, and O.~Pisanti,
  {\it {Bounds on very low reheating scenarios after Planck}},  {\em Phys. Rev.
  D} {\bf 92} (2015), no.~12 123534,
  [\href{http://arxiv.org/abs/1511.00672}{{\tt arXiv:1511.00672}}].

\bibitem{Kolb:1990vq}
E.~W. Kolb and M.~S. Turner, {\em {The Early Universe}}, vol.~69.
\newblock 1990.

\bibitem{Hochberg:2018rjs}
Y.~Hochberg, E.~Kuflik, R.~Mcgehee, H.~Murayama, and K.~Schutz, {\it {Strongly
  interacting massive particles through the axion portal}},  {\em Phys. Rev. D}
  {\bf 98} (2018), no.~11 115031, [\href{http://arxiv.org/abs/1806.10139}{{\tt
  arXiv:1806.10139}}].

\bibitem{Workman:2022ynf}
{\bf Particle Data Group} Collaboration, R.~L. Workman and Others, {\it {Review
  of Particle Physics}},  {\em PTEP} {\bf 2022} (2022) 083C01.

\bibitem{Frangipane:2021rtf}
E.~Frangipane, S.~Gori, and B.~Shakya, {\it {Dark Matter Freeze-In with a Heavy
  Mediator: Beyond the EFT Approach}},
  \href{http://arxiv.org/abs/2110.10711}{{\tt arXiv:2110.10711}}.

\bibitem{Harigaya:2014waa}
K.~Harigaya, M.~Kawasaki, K.~Mukaida, and M.~Yamada, {\it {Dark Matter
  Production in Late Time Reheating}},  {\em Phys. Rev. D} {\bf 89} (2014),
  no.~8 083532, [\href{http://arxiv.org/abs/1402.2846}{{\tt arXiv:1402.2846}}].

\bibitem{Harigaya:2019tzu}
K.~Harigaya, K.~Mukaida, and M.~Yamada, {\it {Dark Matter Production during the
  Thermalization Era}},  {\em JHEP} {\bf 07} (2019) 059,
  [\href{http://arxiv.org/abs/1901.11027}{{\tt arXiv:1901.11027}}].

\bibitem{Bernal:2019mhf}
N.~Bernal, F.~Elahi, C.~Maldonado, and J.~Unwin, {\it {Ultraviolet Freeze-in
  and Non-Standard Cosmologies}},  {\em JCAP} {\bf 11} (2019) 026,
  [\href{http://arxiv.org/abs/1909.07992}{{\tt arXiv:1909.07992}}].

\bibitem{Garcia:2017tuj}
M.~A.~G. Garcia, Y.~Mambrini, K.~A. Olive, and M.~Peloso, {\it {Enhancement of
  the Dark Matter Abundance Before Reheating: Applications to Gravitino Dark
  Matter}},  {\em Phys. Rev. D} {\bf 96} (2017), no.~10 103510,
  [\href{http://arxiv.org/abs/1709.01549}{{\tt arXiv:1709.01549}}].

\bibitem{HotQCD:2018pds}
{\bf HotQCD} Collaboration, A.~Bazavov et~al., {\it {Chiral crossover in QCD at
  zero and non-zero chemical potentials}},  {\em Phys. Lett. B} {\bf 795}
  (2019) 15--21, [\href{http://arxiv.org/abs/1812.08235}{{\tt
  arXiv:1812.08235}}].

\bibitem{Hertel:2018aal}
S.~A. Hertel, A.~Biekert, J.~Lin, V.~Velan, and D.~N. McKinsey, {\it {Direct
  detection of sub-GeV dark matter using a superfluid $^4$He target}},  {\em
  Phys. Rev. D} {\bf 100} (2019), no.~9 092007,
  [\href{http://arxiv.org/abs/1810.06283}{{\tt arXiv:1810.06283}}].

\bibitem{vonKrosigk:2022vnf}
B.~von Krosigk et~al., {\it {DELight: a Direct search Experiment for Light dark
  matter with superfluid helium}},  in {\em {14th International Workshop on the
  Identification of Dark Matter 2022}}, 9, 2022.
\newblock \href{http://arxiv.org/abs/2209.10950}{{\tt arXiv:2209.10950}}.

\bibitem{Bernal:2020qyu}
N.~Bernal, J.~Rubio, and H.~Veerm\"ae, {\it {UV Freeze-in in Starobinsky
  Inflation}},  {\em JCAP} {\bf 10} (2020) 021,
  [\href{http://arxiv.org/abs/2006.02442}{{\tt arXiv:2006.02442}}].

\bibitem{Bernal:2020bfj}
N.~Bernal, J.~Rubio, and H.~Veerm\"ae, {\it {Boosting Ultraviolet Freeze-in in
  NO Models}},  {\em JCAP} {\bf 06} (2020) 047,
  [\href{http://arxiv.org/abs/2004.13706}{{\tt arXiv:2004.13706}}].

\bibitem{Zstats}
P.~N. Bhattiprolu, S.~P. Martin, and J.~D. Wells, ``{\sc Zstats v2.0}
  package.''
  \href{https://github.com/prudhvibhattiprolu/Zstats/}{https://github.com/prudhvibhattiprolu/Zstats/},
  2022.

\bibitem{Bhattiprolu:2022xhm}
P.~N. Bhattiprolu, S.~P. Martin, and J.~D. Wells, {\it {Statistical
  significances and projections for proton decay experiments}},
  \href{http://arxiv.org/abs/2210.07735}{{\tt arXiv:2210.07735}}.

\end{thebibliography}\endgroup

\end{document}